\documentclass[journal]{IEEEtran}

\ifCLASSINFOpdf
\usepackage[pdftex]{graphicx}
\DeclareGraphicsExtensions{.pdf,.jpeg,.png}
\else
\usepackage[dvips]{graphicx}
\fi
\usepackage{subfigure}
\usepackage{epstopdf}
\usepackage[cmex10]{amsmath}
\usepackage{amssymb}
\usepackage{wasysym}
\usepackage{relsize}
\usepackage{algorithm}
\usepackage{algorithmic}
\usepackage{array}
\usepackage{cite}
\usepackage{color}
\usepackage{url}
\usepackage{bm}
\usepackage{enumerate}
\usepackage{epsfig,latexsym}
\usepackage{cite}
\begin{document}
%
\title{Triple-Refined Hybrid-Field Beam Training for  mmWave Extremely Large-Scale MIMO}

\author{Kangjian~Chen,~\IEEEmembership{Student~Member,~IEEE}, Chenhao~Qi,~\IEEEmembership{Senior~Member,~IEEE}, \\ Octavia A. Dobre,~\IEEEmembership{Fellow,~IEEE}  and Geoffrey Ye Li,~\IEEEmembership{Fellow,~IEEE}
	\thanks{This work was supported in part by the National Natural Science Foundation of China under Grants 62071116 and U22B2007, in part  by the  National Key Research and Development Program of China under Grant 2021YFB2900404, in part by the Natural Sciences and Engineering Research Council of Canada (NSERC) through its Discovery program, and in part by the Postgraduate Research \& Practice Innovation Program of Jiangsu Province under Grant KYCX23\_0262. Part of this work has been accepted by IEEE Global Communications Conference, Kuala Lumpur, Malaysia, Dec. 2023~\cite{GlobeCom23CKJ}. (\textit{Corresponding author: Chenhao~Qi})}
	\thanks{Kangjian~Chen and Chenhao~Qi are with the School of Information Science and Engineering, Southeast University, Nanjing 210096, China (e-mail: qch@seu.edu.cn).}
	\thanks{Octavia A. Dobre is with the Faculty of Engineering and Applied Science, Memorial University, St. John’s, NL A1C 5S7, Canada (e-mail: odobre@mun.ca).}
	\thanks{Geoffrey Ye Li is with the Department of Electrical and Electronic Engineering, Imperial College London, SW7 2AZ London, U.K. (e-mail: geoffrey.li@imperial.ac.uk).}
}
\markboth{Accepted by IEEE Transactions on Wireless Communications}
{}
\maketitle

\begin{abstract}
This paper investigates beam training for extremely large-scale multiple-input multiple-output systems. By considering both the near field and far field, a triple-refined hybrid-field beam training scheme is proposed, where high-accuracy estimates of channel parameters are obtained through three steps of progressive beam refinement. First, the hybrid-field beam gain (HFBG)-based first refinement method is developed. Based on the analysis of the HFBG, the first-refinement codebook is designed and the beam training is performed accordingly to narrow down the potential region of the channel path. Then, the maximum likelihood (ML)-based and principle of stationary phase (PSP)-based second refinement methods are developed. By exploiting the measurements of the beam training, the ML is used to estimate the channel parameters. To avoid the high computational complexity of ML, closed-form estimates of the channel parameters are derived according to the PSP. Moreover, the Gaussian approximation (GA)-based third refinement method is developed. The hybrid-field neighboring search is first performed to identify the potential region of the main lobe of the channel steering vector. Afterwards, by applying the GA, a least-squares estimator is developed to obtain the high-accuracy channel parameter estimation. Simulation results verify the effectiveness of the proposed scheme. 
\end{abstract}

\section{Introduction}
Millimeter wave (mmWave) that reserves wide spectrum resources is a promising technology for achieving high data rates. Thanks to its small wavelength, the space-limited  base station (BS) can accommodate a large number of antennas to  enhance the spectral efficiency by implementing the massive multiple-input multiple-output (MIMO) ~\cite{WC19BE,Tcom22CMHGJJ,TWC22QZK}. The perfect match between the abundant spectrum resources provided by mmWave and the high spectral efficiency  enabled by massive MIMO has led to the boom of  mmWave massive MIMO~\cite{SciChinaChenhao2021}. 
 
Recently, extremely large-scale MIMO (XL-MIMO) with far more antennas than existing massive MIMO has been  developed to further improve the spectral efficiency via ultrahigh-gain beamforming~\cite{CM23CMY,Tcom22CMH,Tcom23LY}.  Attracted by the tremendous potentials, researchers move on to develop XL-MIMO based on the well-explored massive MIMO. However, extending the works in massive MIMO to XL-MIMO is not straightforward. Due to the much larger array aperture, the channel model of the latter differs greatly from that of the former. Generally, the radiation fields of the electromagnetic waves can be divided into the near field and the far field according to the distance between the BS and the radiation source~\cite{APM17SKT, TWC22LHQ}. In the far field, the channel characteristics conform to the planar-wave model, and the phase differences among antennas can be reasonably approximated as a linear function of the antenna indices~\cite{Tcom22CMH}. In the near field, the channel characteristics cannot be accurately modeled using the planar-wave model. Instead, they must be characterized by the spherical-wave model, where the phase differences among antennas are expressed as a nonlinear function of the antenna indices~\cite{Tcom22CMH}. For conventional massive MIMO, where the array aperture is relatively small, the BS coverage predominantly falls within the far field.  However, in the emerging XL-MIMO, the substantial expansion of the array aperture results in a significant portion of the BS coverage lying within the near field.  In this context, XL-MIMO communications will operate within the coexistence of the far and near fields. In addition, the difference in channel models renders conventional far-field techniques unsuitable for the near field. Therefore, in the realm of XL-MIMO communications, the focus should be shifted towards techniques that can effectively adapt to the hybrid field, including the near and far fields.
 
One pivotal issue in wireless communications  is the channel state information (CSI) acquisition~\cite{WC21CZ,OPCS23NBY,TWC22NBY}. Due to the large propagation attenuation of mmWave, beam training that can achieve high beamforming gain is preferred~\cite{TWC19SXY}. Generally, beam training can be categorized into beam sweeping and multi-stage beam training. Although beam sweeping  can effectively combat noise by using narrow beams, its success is founded on the intensive beam training. In contrast, multi-stage beam training that explores and refines the CSI stage by stage can compare favorably with the beam sweeping while requiring much lower training overhead~\cite{JSAC17LCS}.  A special case of the multi-stage beam training is the hierarchical beam training (HBT)~\cite{JSAC09WJY,ZhenyuXiao2018,TWC20QCH,TVT21Ningboyu,IOT23NBY}. By comparing the received powers of the codewords in the predefined hierarchical codebook layer by layer,  the HBT can gradually narrow down the candidate sets of channel paths, leading to the increasing beamforming gain. Different from the HBT that depends on the hierarchical codebook and mainly exploits the power information of the received signals, the general multi-stage beam training flexibly divides the whole training process into several stages and extracts CSI from the former stage via signal processing techniques to assist the beam training of the latter stage~\cite{TWC19LM,JSAC19CSE,TWC17ZDL,TVT23ZA}. In \cite{TWC19LM}, an optimized two-stage search algorithm is developed, where the second stage only trains the beam candidates derived from the first stage to accomplish beam alignment.  In \cite{JSAC19CSE}, an adaptive and sequential beam alignment algorithm is proposed, where the codeword for the beam training in the lower layer is determined by the posterior of the channel angle derived in the upper layer. In~\cite{TWC17ZDL}, based on the beam sweeping results, an auxiliary beam pair method is proposed to obtain the high-accuracy estimates of channel angles by comparing the powers of received signals. In~\cite{TVT23ZA}, to further refine the result of beam sweeping, the far-field array gain is approximated as a Gaussian function and the channel angles are estimated based on additional channel~tests.

The fundamental changes in the radiation fields for XL-MIMO communication invalidate the conventional far-field beam training methods in~\cite{TWC22NBY,TWC19SXY,JSAC17LCS,ZhenyuXiao2018,TWC20QCH,TVT21Ningboyu,JSAC09WJY,IOT23NBY,TWC19LM,JSAC19CSE,TWC17ZDL,TVT23ZA}. In addition, the coexistence of the near and far fields appeals for novel hybrid-field beam training methods for XL-MIMO communications. One straightforward method for hybrid field beam training is the hybrid-field beam sweeping (HFBS)~\cite{ICCC2022CKJ}, which explores the far field via beam scanning and the near field via beam focusing. Nevertheless, the HFBS exhaustively tests every codeword in the predefined hybrid-field codebook and needs an extensive amount of beam training. To reduce the overhead of HFBS, a two-phase beam training (TPBT) method is proposed in~\cite{WCL22ZYP}, where the first phase determines the candidate channel angles via far-field beam sweeping while the second phase finds the channel distance based on shortlisted candidate angles in the first phase.  However, the overhead of the TPBT is still high due to the employment of the far-field beam sweeping in the first phase. To further reduce the overhead of TPBT, a hierarchical codebook is designed and a chirp-based HBT (CHBT) method is proposed for the XL-MIMO in \cite{TWC23SX}.  It is worth noting that the imperfection in the beam patterns of codewords in the hierarchical codebook could lead to degradation in beam training performance. For the distance-based HBT (DHBT) method in~\cite{CC22WXH}, the codebook is designed by equally sampling the space in distance. However, the neglect of the polar-domain sparsity leads to  a deterioration in the training performance of DHBT. For the  two-stage HBT (TSHBT) in \cite{TVT23WCY}, the training procedure is divided into two stages, where the first stage only refines the channel angle with wide beams while the second stage simultaneously refines the channel angle and distance with beam focusing. However, a large number of antennas need to be deactivated to form wide beams in the first stage.  Owing to the powerful feature extraction ability of neural networks, deep learning has also been exploited for hybrid-field beam training~\cite{CL23JGL,CL23LW,Tcom23LW}.  However, the substantial computational overhead and resource-intensive characteristics of deep learning present significant barriers to its widespread implementation in XL-MIMO systems. Furthermore, the beam training methods in \cite{ICCC2022CKJ,WCL22ZYP,TWC23SX,CC22WXH,TVT23WCY,CL23JGL,CL23LW} construct codebooks based on the quantized samples, which not only deteriorates the beamforming gain but also  increases errors in near-field localization. While numerous studies have explored high-accuracy near-field sensing and localization~\cite{TSP10LJL,CL23_WZL,TSP19FB}, they often rely on complicated hardware configurations or necessitate extensive computational resources, which is incompatible with the concise beam training framework. One simple method to improve the accuracy of hybrid-field beam training is the beam refinement~\cite{TWC23CKJ}. However, the beam refinement method in~\cite{TWC23CKJ} is specially tailored for the partially-connected hybrid beamforming structure. In the context of XL-MIMO, beam refinement methods for the fully-connected hybrid beamforming structure are missing in the existing works.  



In this paper, we investigate hybrid-field beam training for XL-MIMO systems. We aim to achieve high-accuracy beam alignment for XL-MIMO systems  with low overhead and low computational complexity, which is beyond the scopes of the existing works in \cite{ICCC2022CKJ,WCL22ZYP,TWC23SX,CC22WXH,TVT23WCY,CL23LW,Tcom23LW,TSP10LJL,CL23_WZL,TSP19FB,TWC23CKJ,CL23JGL}. Inspired by the idea of the multi-stage beam training, we propose to acquire the CSI of XL-MIMO systems in a progressive refinement way. To obtain the high-accuracy CSI with low overhead, efficient refinement methods and effective progressive strategies  are devised. The main contributions of this paper are summarized as follows.

\begin{itemize}
	\item By considering both the near field and far field, we develop a triple-refined hybrid-field beam training (THBT) scheme, where high-accuracy estimates of channel parameters are obtained through three steps of progressive refinement. Benefiting from the devised refinement methods and progressive strategies, the proposed THBT can accomplish the beam alignment for XL-MIMO systems with substantially reduced training overhead. 
	
 	\item We develop the hybrid-field beam gain (HFBG)-based first refinement method. By analyzing the HFBG of the channel steering vector for the XL-MIMO systems, we design the first-refinement codebook that satisfies the presented two design criteria and  perform the beam training accordingly to determine the potential region of the channel path.
 	
 	\item We then exploit the maximum likelihood (ML) and principle of stationary phase (PSP) to develop the second refinement methods. Based on the property of the HFBG, we first design the second-refinement codebook, where the beam coverage of each codeword contains the potential region from the first refinement. Then, by exploiting the measurements of the beam training, the ML is used to estimate the channel parameters. To avoid the high computational complexity of ML, the phase of the hybrid-field beam gain is analytically expressed according to the PSP, and closed-form estimates of the channel parameters are derived.
 	
 	\item We develop the Gaussian approximation (GA)-based third refinement method. First, we perform the hybrid-field neighboring search to identify the potential region of the main lobe of the channel steering vector. Inspired by the similarities between the Taylor series of the Gaussian function and that of the HFBG, we approximate the main lobe of the HFBG as a two-dimensional Gaussian function. Then, by sampling the surrogate distance and angle uniformly within the potential region, we design the third-refinement codebook and perform the beam training accordingly. Based on the measurements of the beam training, a least-squares (LS) estimator is developed to obtain the high-accuracy channel parameter~estimation.
\end{itemize}

The rest of this paper is organized as follows. The model of the XL-MIMO system is introduced in Section~\ref{SystemModel}. The codebook design and beam training for the first, second, and third refinement are presented in Sections~\ref{FLBT},~\ref{SLBT},~and \ref{SecThirdLayer}, respectively. The proposed methods are evaluated in Section~\ref{SimulationResults}, and the paper is concluded in Section~\ref{Conclusion}.

The notations are defined as follows. Symbols for matrices (upper case) and vectors (lower case) are in boldface. $(\cdot)^*$, $(\cdot)^{\rm T}$, and $(\cdot)^{\rm H}$ denote the conjugate, transpose, and conjugate transpose (Hermitian), respectively.  $[\boldsymbol{a}]_{n}$,  $\left[ \boldsymbol{A} \right] _{m,:}$, $\left[ \boldsymbol{A} \right] _{:,n}$, and $\left[ \boldsymbol{A} \right] _{m,n}$ denote the $n$th entry of vector $\boldsymbol{a}$, the $m$th row of matrix $\boldsymbol{A}$, the $n$th column of matrix $\boldsymbol{A}$, and the entry on the $m$th row and the $n$th column of matrix $\boldsymbol{A}$, respectively. $\{\boldsymbol{\mathcal{C}}\}_n$ is the $n$th element of the set $\boldsymbol{\mathcal{C}}$. In addition, $j$,~$|\cdot|$,~$\mathbb{C}$,~and $\mathcal{C}\mathcal{N}$ denote the square root of $-1$, the absolute value of a scalar, the set of the complex number and the complex Gaussian distribution. Moreover, $f'(\cdot)$ and $f''(\cdot)$ represent the first-order and the second-order derivative of $f(\cdot)$, respectively. $\mbox{ln}(\cdot)$ denotes the  natural logarithm of a number.

\section{System Model}\label{SystemModel}
\begin{figure}[!t]
	\centering
	\includegraphics[width=70mm]{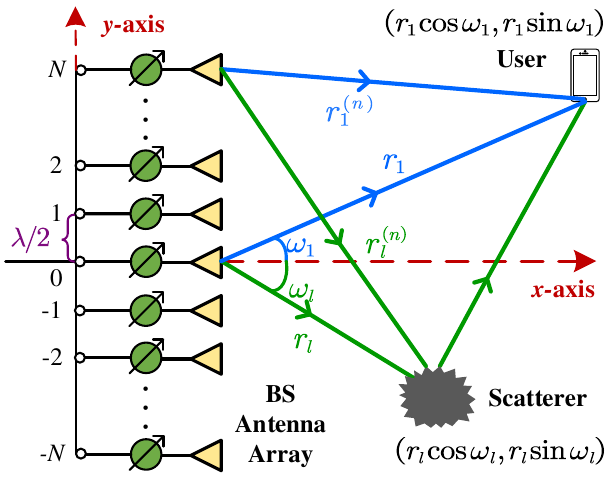}
	\caption{Illustration of the system model.}
	\label{MultipathChannelModel}
\end{figure}

As shown in Fig.~\ref{MultipathChannelModel}, we consider the downlink beam training between the BS and the user, where a half-wavelength-interval array with $N_{\rm t}$ antennas is equipped at the BS while a single-antenna transceiver is adopted at the user. To ease the notation, we assume that $N_{\rm t}$ is an odd number, and $N_{\rm t} = 2N+1$. With the hybrid beamforming structure, the $N_{\rm t}$ antennas are connected to $N_{\rm RF}$ radio frequency (RF) chains through a phase shifter network. In this work, the beam training is performed based on the analog beamforming, and hence, we only focus on one of the multiple RF chains for simplicity.  Then the $s$th received signal, for $s=1,2,\ldots, S$, can be expressed~as 
\begin{align}\label{systemmodel}
	y_s = \boldsymbol{h}^{\rm H}\boldsymbol{f}_sx_s + \eta
\end{align}
where $\boldsymbol{h}\in\mathbb{C}^{N_{\rm t}}$, $\boldsymbol{f}_s\in \mathbb{C}^{N_{\rm t}}$, and $x_s$ denote the channel between the BS and the user, the analog beamformer at the BS,  and the transmit signal, respectively. In addition, $\eta\sim\mathcal{CN}(0,\varrho^2)$ denotes the additive white Gaussian noise.

To depict the channel, we first establish a Cartesian coordinate, which sets the center, the normal direction, and the tangent direction of the  antenna array as the origin, the x-axis, and the y-axis, respectively. Then the coordinate of the $n$th antenna can be expressed as $(0, n\lambda/2)$ for $n\in\boldsymbol{\mathcal{I}}$, where $\boldsymbol{\mathcal{I}}\triangleq\{-N,\cdots,0,\cdots,N\}$ and  $\lambda$ denotes the wavelength. From Fig.~\ref{MultipathChannelModel}, the coordinate of the radiation source at the $l$th path is $(r_l\cos\omega_l,r_l\sin\omega_l)$, for $l=1,2,\cdots,L$, where $L$ denotes the number of paths, $r_l$ is the distance between the origin and the $l$th radiation source, and $\omega_l\in[-\pi/2,\pi/2]$ represents the angle of the $l$th radiation source relative to the x-axis. The distance between the $l$th radiation source and the $n$th antenna is calculated as 
\begin{equation}\label{RangeRln}
	r_l^{(n)} = \sqrt{r_l^2 + n^2\lambda^2/4 - nr_l\Omega_l\lambda}
\end{equation}
where $\Omega_l\triangleq\sin\omega_l \in[-1,1]$. Then, the channel steering vector between the BS and the user can be expressed as
\begin{equation}\label{ChannelModel}
	\boldsymbol{h} = \sum_{l=1}^{L}g_l\boldsymbol{\alpha}(\Omega_l,r_l)
\end{equation}
where $g_l$ denotes the channel gain of the $l$th path and the channel steering vector $\boldsymbol{\alpha}(\cdot)$ is defined as
\begin{equation}\label{steeringvector}
	\boldsymbol{\alpha}(\Omega_l,r_l)\! =\! \frac{1}{\sqrt{N_{\rm t}}}\!\left[e^{-j\frac{2\pi}{\lambda}(r_l^{(-N)}\!-\!r_l)},\!\ldots\!,e^{-j\frac{2\pi}{\lambda} (r_l^{(N)}-r_l)}\right]^{\rm T}.
\end{equation}
Note that the channel steering vector in \eqref{steeringvector} adapts to both the near and far fields. To ease the notation, we omit the subscript ``$l$"  and focus on processing the path of interest. Then, $r^{(n)}$ can be simplified~as
\begin{align}\label{Approx2D}
	r^{(n)} \approx r - n\Omega\lambda/2 + \frac{n^2\lambda^2(1-\Omega^2)}{8r}
\end{align}
according to $\sqrt{1+\epsilon}\approx1+\frac{1}{2}\epsilon - \frac{1}{8}\epsilon^2$, which is verified to be accurate if $r^{(n)}\ge 0.5\sqrt{N^3\lambda^2}$~\cite{Tcom22CMH,APM17SKT}. In fact, $0.5\sqrt{N^3\lambda^2}$ is quite small compared to the coverage of the BS. For example, if $N = 128$ and $\lambda = 0.005$~m, we have $0.5\sqrt{N^3\lambda^2} \approx 3.6$~m, which is much smaller than the typical coverage of the BS. Therefore, in this work, we focus on the radiation field with $r^{(n)}\ge 0.5\sqrt{N^3\lambda^2}$. Substituting \eqref{Approx2D} into \eqref{steeringvector}, we have 
\begin{align}
	[\boldsymbol{\alpha}( \Omega,r)]_n &\approx \frac{1}{\sqrt{N_{\rm t}}} e^{j\pi\left( \Omega n - \frac{\lambda(1-\Omega^2)}{4r}n^2\right)}
\end{align}	
for $n\in\boldsymbol{\mathcal{I}}$. We define 
\begin{align}\label{Definitionk}
 b \triangleq \frac{\lambda(1- \Omega^2)}{4r}
\end{align}
which is referred to as the ``surrogate distance" in this paper. Then \eqref{steeringvector} can be approximated as 
\begin{align}\label{steeringvectorapprox}
	\boldsymbol{\alpha}(\Omega,r)\!&\approx\frac{1}{\sqrt{N_{\rm t}}}\left[e^{j\pi(-\Omega N-b N^2)},\ldots,e^{j\pi(\Omega N-b N^2)}\right]^{\rm T}\nonumber \\
	&\triangleq \boldsymbol{\gamma}(\Omega,b).
\end{align}

We then propose the THBT scheme to obtain high-accuracy  channel parameter estimation through three steps of progressive refinement.

\section{HFBG-based First Refinement Method}\label{FLBT}
In this section, we first analyze the HFBG of the channel steering vector for XL-MIMO systems. Based on the analysis, we develop the HFBG-based first refinement method, where the first-refinement codebook is designed and the beam training is performed accordingly to determine the potential region of the channel path.

\subsection{Hybrid-Field Beam Gain}
For an arbitrary steering vector, $\boldsymbol{u}\triangleq\boldsymbol{\gamma}(\Theta,k)$, we define its HFBG as
\begin{align}\label{BeamGain}
	G(\boldsymbol{u},\Omega,b) &\triangleq N_{\rm t}\boldsymbol{\gamma}(\Omega,b)^{\rm H}\boldsymbol{u} \nonumber \\
	&= \sum_{n=-N}^{N}e^{j\pi((\Theta-\Omega)n + (b-k)n^2)}\nonumber \\
	&\overset{\rm (a)}{\approx}\!\int_{-N-1/2}^{N+1/2}e^{j\pi((\Theta-\Omega)z + (b-k)z^2)}dz \nonumber \\
	&=\int_{-\infty}^{\infty}\!U(z)e^{j\pi((\Theta-\Omega)z + (b-k)z^2)}dz
\end{align}
where
\begin{align}\label{Uz}
	U(z) = \left\{ \begin{array}{ll}
		1, & -N-1/2\leq z \leq N+1/2\\
		0, & \mbox{others}.
	\end{array} \right.
\end{align}
In \eqref{BeamGain}, we approximate the summation as the integral in $\rm (a)$. Due to the quadratic phase structure of the integrand, it is hard to obtain the closed-form solution of \eqref{BeamGain}. Alternatively, a simple but effective approximation based on the PSP is widely adopted~\cite{AMFI,AMMSE,ASF}. We define 
\begin{align}
	J(z,\Omega,b) \triangleq \pi((\Theta-\Omega)z + (b-k)z^2)
\end{align}
and denote $z_0$ as the stationary phase. According to the PSP, the differential of $J(z,\Omega,b)$ at the stationary phase satisfies the condition that $J'(z_0,\Omega,b) = 0$. Therefore, we can obtain
\begin{align}\label{eqz0}
	z_0 = \frac{\Omega-\Theta}{2(b-k)}.
\end{align}
Then, based on the PSP,  \eqref{BeamGain} can be approximated as 
\begin{align}\label{PSPApp}
	G(\boldsymbol{u},\Omega,b) &\approx \sqrt{\frac{-2\pi}{J''(z_0,\Omega,b)}}e^{-j\pi/4}U(z_0)e^{jJ(z_0,\Omega,b)}\nonumber \\
	&=\frac{e^{-j\pi/4}}{\sqrt{k-b}}U(z_0)e^{j\pi\frac{(\Omega-\Theta)^2}{4(k-b)}} \nonumber \\
	&=\left\{ \begin{array}{ll}
		\!\frac{e^{-j\pi/4}}{\sqrt{k-b}}\!e^{j\pi\frac{(\Omega-\Theta)^2}{4(k-b)}},&\!-\!\frac{N_{\rm t}}{2}\!\le\!z_0\!\le\!\frac{N_{\rm t}}{2}\\
		0, &\!\mbox{others}.
	\end{array} \right.
\end{align}
By taking the absolute value of $G(\boldsymbol{u},\Omega,b)$, we have 
\begin{align}\label{absolutebeamgain}
|G(\boldsymbol{u},\Omega,b)|\approx\left\{ \begin{array}{ll}
		\!\frac{1}{\sqrt{|b-k|}}, &-\frac{N_{\rm t}}{2}\le z_0 \le \frac{N_{\rm t}}{2}\\
		0, &\!\mbox{others}.
	\end{array} \right.
\end{align}
Combining \eqref{eqz0} and \eqref{absolutebeamgain}, we can express the beam coverage of $\boldsymbol{u}$~as 
\begin{align}\label{coverage}
	\boldsymbol{\mathcal{B}}(\boldsymbol{u}) = \left\{(\Omega,b)\bigg| \frac{|\Omega-\Theta|}{|b-k|}\le N_{\rm t}\right\}.
\end{align}
From \eqref{coverage}, the angle coverage of $\boldsymbol{u}$ for a fixed $b$ is 
\begin{align}\label{anglecoverage}
	\Theta-N_{\rm t}|b-k|\le\Omega\le \Theta + N_{\rm t}|b-k|
\end{align}
which indicates that the beam coverage is symmetric about angle $\Theta$ and the angle coverage width (ACW) for a fixed $b$ is $2N_{\rm t}|k-b|$.
\begin{figure}[!t]
	\begin{center}
		\includegraphics[width=80mm]{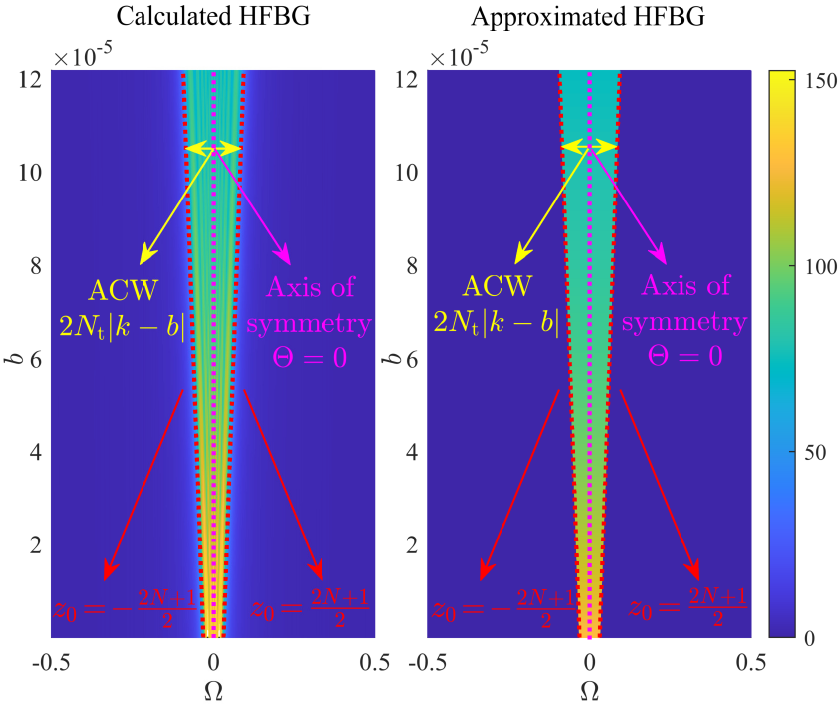}
	\end{center}
	\caption{Comparison of the calculated and the approximated HFBGs.}
	\label{figPSP}
\end{figure}

In Fig.~\ref{figPSP}, we illustrate the calculated and the approximated HFBGs of~$\boldsymbol{u}$, where we set $N_{\rm t} = 513$, $N = 256$, $\Theta = 0$ and $k = -6.09\times10^{-5}$. From the figure, the calculated HFBG and the approximated HFBG share similarities in several aspects, including the boundary, the ACW and the axis of the symmetry, which indicates that the PSP can provide a good approximation for the HFBG.

\subsection{First-Refinement Codebook Design and Beam Training}


Suppose the initial potential region of the interested channel path is 
\begin{align}\label{initlalInterstedRegion}
	\boldsymbol{\Phi}_1 = \big\{(\Omega,b)\big|-\overline{\Omega}\le\Omega\le\overline{\Omega}, 0\le b\le\overline{b}\big\}
\end{align} 
where $\overline{\Omega}$ and $\overline{b}$ denote boundaries of $\Omega$ and $b$, respectively. Note that in \eqref{initlalInterstedRegion} we assume the angle of the potential region is symmetric about zero for simplicity. However, more complicated cases can also be developed by exploiting the phase-shift invariance property of the channel steering vector~\cite{TWC23SX}.  In addition, $b$ is lower bounded by zero due to its definition in~\eqref{Definitionk}.

Generally, the beam training focuses on narrowing down the potential region by testing the codewords in the predefined codebook. Then, directional beams with high beamforming gains can be formed to combat the noise. To narrow down the potential region effectively and efficiently, the codeword design commonly considers the following two criteria~\cite{TSP20CKJ}: 1) The beam coverages of the codewords do not overlap with each other. 2) The potential region should be fully covered by the union of the beam coverage of the codewords. In the following, by considering the aforementioned two criteria, the first-refinement codebook is designed accordingly.

\begin{figure}[!t]
	\begin{center}
		\includegraphics[width=80mm]{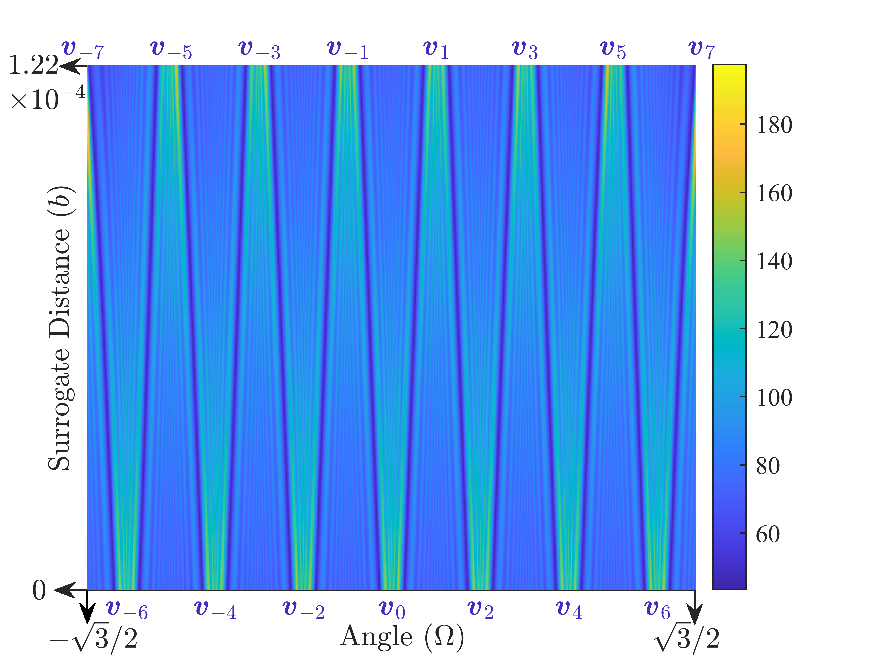}
	\end{center}
	\caption{Illustration of the beam gain of $\boldsymbol{v}_m$ for $m = -M_1,\cdots,0,\cdots M_1$.}
	\label{IlluCodewords}
\end{figure}

Denote the first-refinement codebook as
\begin{align}\label{flcodebook}
	 \boldsymbol{\mathcal{C}}_1 = \{\boldsymbol{v}_{-M_1},\cdots,\boldsymbol{v}_{0},\cdots,\boldsymbol{v}_{M_1}\}
\end{align} 
where $2M_1 +1$ is the number of codewords.  We design the $m$th codeword in $\boldsymbol{\mathcal{C}}_1$ as
\begin{align}\label{wc1}
	\boldsymbol{v}_m = \boldsymbol{\gamma}\big(m\overline{\Theta}_1,\overline{k}_1\big)
\end{align}
for $m =-M_1,-M_1+1,\cdots, M_1$, where 
\begin{align}\label{widetildek2}
	\overline{k}_1 =		\left\{ \begin{array}{ll}
		\widetilde{k}_1, &m~\mbox{is even,} \\
		\overline{b} -\widetilde{k}_1, &m~\mbox{is odd,}
	\end{array} \right.
\end{align}
$\widetilde{k}_1<0$ is the introduced parameter to control the beam coverage  of the designed codewords, and $\overline{\Theta}_1 \triangleq \big(\overline{b} - 2\widetilde{k}_1\big)N_{\rm t}$ denotes the angle deviation between adjacent codewords. By substituting \eqref{wc1} and \eqref{widetildek2} into \eqref{coverage},  it is convenient to verify that the beam coverages of the designed codewords fit closely without overlapping, which indicates that the first criterion is satisfied. In addition, we set only two values for $\overline{k}_1$ in \eqref{widetildek2} so that the designed codewords have similar  beam coverages and can provide comparable beamforming gains for every position in the space. To satisfy the second criterion, the potential region in \eqref{initlalInterstedRegion} should be fully covered by the $2M_1 +1$ codewords in \eqref{flcodebook}. Since both the potential region and the beam coverage of the designed codewords are symmetric, we only focus on the region with $\Omega\ge 0$. From \eqref{anglecoverage}, the right angle boundary of $\boldsymbol{v}_{M_1}$ is $M_1\overline{\Theta}_1 + N_{\rm t}|b-\overline{k}_1|$, where $0\le b \le\overline{b}$. Then according to \eqref{initlalInterstedRegion} and the second criterion, we have 
\begin{align}\label{anglecoverageM1}
& \min_{0\le b\le\overline{b} } \ \ M_1\overline{\Theta}_1 + N_{\rm t}|b-\overline{k}_1| > \overline{\Omega}.
\end{align}
From \eqref{anglecoverageM1}, we have
\begin{align}\label{anglecoverageM1m}
 M_1 > ( \overline{\Omega} +  N_{\rm t}\widetilde{k}_1)/\overline{\Theta}_1
\end{align}
which provides guidance for the design of $M_1$.

Then, we perform the beam training based on $\boldsymbol{\mathcal{C}}_1$ to narrow down the potential region. When testing the $m$th codeword in $\boldsymbol{\mathcal{C}}_1$, the  signal received by the user can be expressed as
\begin{align}\label{firstlayerbt}
	y_m = \boldsymbol{h}^{\rm H}\boldsymbol{v}_m + \eta
\end{align}
for $m = -M_1,\cdots,0,\cdots,M_1$, where we set the transmit signal as ``1". The result of the beam training can be expressed~as 
\begin{align}\label{firstlayerbtr}
	\overline{m}  = \arg\max_{m = -M_1,\cdots,0,\cdots, M_1} |y_m|.
\end{align} 
From \eqref{firstlayerbtr}, the channel path locates in the beam coverage of $\boldsymbol{v}_{\overline{m}}$. Therefore, we can update the potential region in \eqref{initlalInterstedRegion} as 
\begin{align}\label{initlalInterstedRegion2}
	\widetilde{\boldsymbol{\Phi}}_1 = \left\{(\Omega,b)\bigg|0\le b\le\overline{b},\frac{|\Omega-\overline{m}\overline{\Theta}_1|}{|b-\overline{k}_1|}\le N_{\rm t}\right\},
\end{align}
where $\overline{k}_1$ can be determined by substituting $\overline{m}$ into \eqref{widetildek2}.

\begin{algorithm}[!t]
	\caption{HFBG-based First Refinement Methods}
	\label{Alg1}
	\begin{algorithmic}[1]
		\STATE \textbf{Input:} $N$, $N_{\rm t}$, $\lambda$, $\boldsymbol{h}$, $\overline{\Omega}$, $\overline{b}$, and $\widetilde{k}_1$
		\STATE Obtain $M_1$ via \eqref{anglecoverageM1m}.
		\STATE Obtain  $\boldsymbol{\mathcal{C}}_1$ via \eqref{flcodebook}.
		\STATE Obtain $y_m$ for $m = -M_1,\cdots,0,\cdots,M_1$ via \eqref{firstlayerbt}.
		\STATE Obtain $\overline{m}$ via \eqref{firstlayerbtr}.
		\STATE Obtain $\widetilde{\boldsymbol{\Phi}}_1$ via \eqref{initlalInterstedRegion2}.
		\STATE \textbf{Output:} $\overline{m}$ and $\widetilde{\boldsymbol{\Phi}}_1$.
	\end{algorithmic}
\end{algorithm}

Now, we provide an example for the  first-refinement codebook in \textbf{Design Example 1}.

\textbf{Design Example 1:} Consider an XL-MIMO system with $N = 256$, $N_{\rm t} = 513$, $\lambda = 0.005$~m, $\widetilde{k}_1 = -6.09\times10^{-5}$, $\overline{b} = 1.22\times 10^{-4}$ and $\overline{\Omega} = \sqrt{3}/2$. From \eqref{anglecoverageM1m}, we can obtain $M_1\ge 6.68$. Since $M_1$ must be an integer, we set $M_1 = 7$, leading to totally $15$ codewords in $\boldsymbol{\mathcal{C}}_1$. According to \eqref{wc1}, $\boldsymbol{v}_m$ for $m = -M_1,\cdots,0,\cdots, M_1$ can be designed. In Fig.~\ref{IlluCodewords}, we illustrate the calculated HFBG of $\boldsymbol{v}_m$ within the potential region $\boldsymbol{\Phi}_1$. From the figure, the potential region can be fully covered by the codewords in $\boldsymbol{\mathcal{C}}_1$ and there is little overlap between the beam coverages of two different codewords in $\boldsymbol{\mathcal{C}}_1$, which indicates that the two aforementioned design criteria are well satisfied.
 
Finally, we summarize the HFBG-based first refinement method in \textbf{Algorithm~\ref{Alg1}}.

\section{ML-based and PSP-based Second Refinement Methods}\label{SLBT}
In this section, by exploiting the ML and PSP, we develop the second refinement methods to further improve the accuracy of channel parameter estimation.

Note that the beam training is more likely to fail when channel paths locate in the transition zone of the codeword due to the similar beam gains of adjacent codewords~\cite{JJZhang2017}. Therefore, we extend the potential region in \eqref{initlalInterstedRegion2} as 
\begin{align}\label{initlalInterstedRegion21}
	\boldsymbol{\Phi}_2  = \left\{(\Omega,b)\bigg|0\le b\le\overline{b},\frac{|\Omega-\overline{m}\overline{\Theta}_1|}{|b-\overline{k}_1|+1/N_{\rm t}^2}\le N_{\rm t}\right\}
\end{align} 
to include the transition zone, where the width of the transition zone is typically set as $1/N_{\rm t}$~\cite{CL19CKJ}. Then we perform the beam training within $\boldsymbol{\Phi}_2$.

We denote the second-refinement codebook as 
\begin{align}\label{secondlayercodebook}
	\boldsymbol{\mathcal{C}}_2 = \{\boldsymbol{w}_{-M_2},\cdots,\boldsymbol{w}_0,\cdots,\boldsymbol{w}_{M_2}\}
\end{align}
where $2M_2+1$ is the number of codewords. We design the $m$th codeword in \eqref{secondlayercodebook} as 
\begin{align}\label{secondlayercodebookcodeword}
	\boldsymbol{w}_{m} = \boldsymbol{\gamma}(\widetilde{\Theta}_m,\overline{k}_2),~	\widetilde{\Theta}_m = \overline{m} \overline{\Theta}_1 + m\overline{\Theta}_2
\end{align}
for $m=-M_2,\cdots,0,\cdots,M_2$, where $\overline{k}_2$ and $\overline{\Theta}_2>0$ are introduced parameters to control the beam coverage and the angle deviation of the codewords, respectively. According to \eqref{coverage}, the beam coverage of the $m$th codeword can be expressed as 
\begin{align}\label{initlalInterstedRegion22}
	\boldsymbol{\Psi}_m = \left\{(\Omega,b)\bigg|0\le b\le\overline{b},\frac{|\Omega-\widetilde{\Theta}_m|}{|b-\overline{k}_2|}\le N_{\rm t}\right\}.
\end{align} 
To provide high beam gain during beam training, the beam coverage of each codeword in $\boldsymbol{\mathcal{C}}_2$ should contain $\boldsymbol{\Phi}_2$. Based on \eqref{initlalInterstedRegion21} and \eqref{initlalInterstedRegion22}, $\overline{k}_2$ should satisfy  
\begin{align}\label{coverageofk2}
\frac{1}{N_{\rm t}} + N_{\rm t}|b-\overline{k}_1| \le N_{\rm t}|b-\overline{k}_2| + m\overline{\Theta}_2.
\end{align}
From \eqref{coverageofk2}, we have 
\begin{align}\label{coverageofk2_2}
	\left\{ \begin{array}{ll}
		\overline{k}_2 \le  \overline{k}_1 - M_2\overline{\Theta}_2/N_{\rm t} - 1/N_{\rm t}^2, &\overline{m}~\mbox{is even} \\
		\overline{k}_2 \ge  \overline{k}_1 + M_2\overline{\Theta}_2/N_{\rm t}+ 1/N_{\rm t}^2, &\overline{m}~\mbox{is odd}.
	\end{array} \right.
\end{align}
Then, we perform the beam training based on $\boldsymbol{\mathcal{C}}_2$. When testing the $m$th codeword in $\boldsymbol{\mathcal{C}}_2$, the signal received by the user can be expressed as
\begin{align}\label{secondlayerbt}
	\overline{y}_m &= \boldsymbol{h}^{\rm H}\boldsymbol{w}_{m} + \eta \nonumber \\
	& \overset{\rm (a)}{\approx} \boldsymbol{h}^{\rm H}\boldsymbol{\gamma}(\widetilde{\Theta}_m,\overline{k}_2) \nonumber \\
	& \overset{\rm (b)}{\approx} g^{*}\boldsymbol{\gamma}(\Omega,b)^{\rm H}\boldsymbol{\gamma}(\widetilde{\Theta}_m,\overline{k}_2)
\end{align}
where we omit the noise term in $\rm (a)$ and the effects of the non-line-of-sight (NLoS) paths in $\rm (b)$. 

\begin{figure*}[!t]
	\centering
	\subfigure[The coverage of $\widetilde{\boldsymbol{\Phi}}_1$, $\boldsymbol{\Phi}_2$ and $\boldsymbol{w}_m$.]
	{
		\begin{minipage}[b]{.37\linewidth}
			\centering
			\includegraphics[scale=0.51]{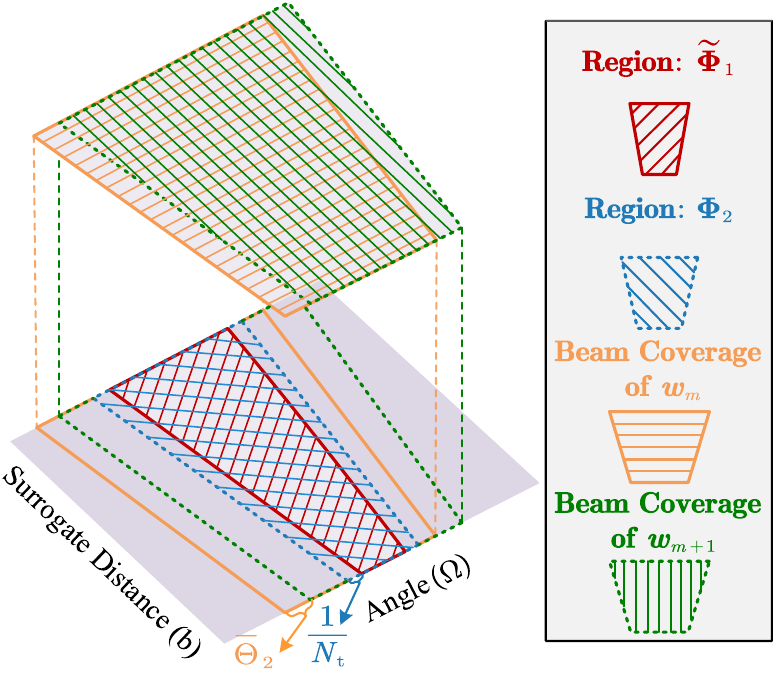}
		\end{minipage}
		\label{Coverage}
	}
	\subfigure[The phase unwrapping and the PSP approximation.]
	{
		\begin{minipage}[b]{.37\linewidth}
			\centering
			\includegraphics[scale=0.51]{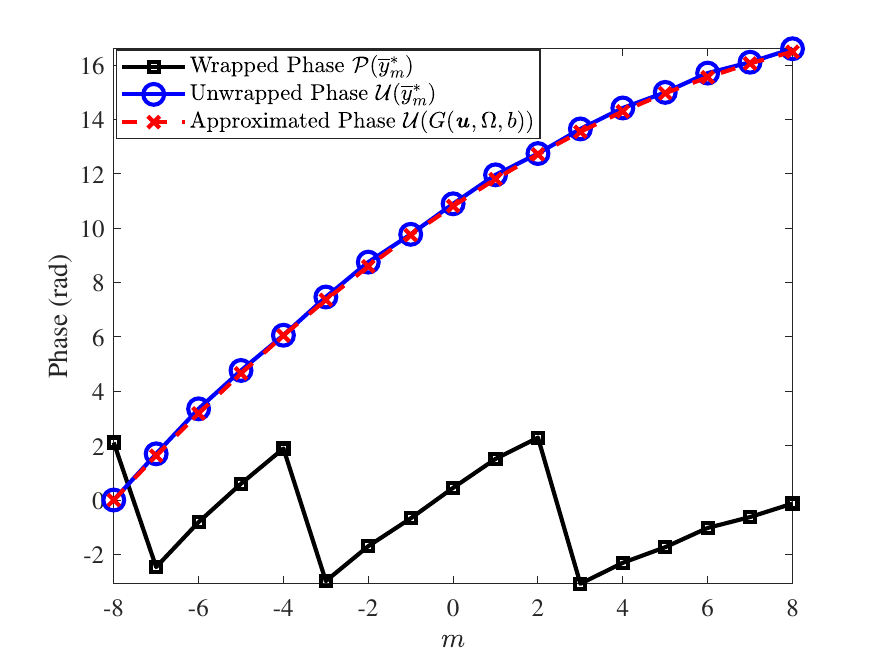}
		\end{minipage}
		\label{PhaseUnwarp}
	}
	\subfigure[Parameters Estimation.]
	{
		\begin{minipage}[b]{.2\linewidth}
			\centering
			\includegraphics[scale=0.51]{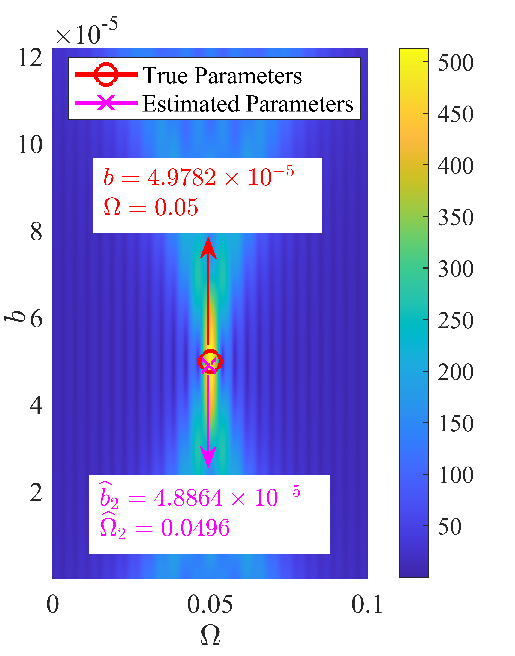}
		\end{minipage}
		\label{ParametersEstimation}
	}
	\caption{Illustration of the second-refinement codebook and beam training.}
	\label{FigCT}
	\vspace{-0.1cm}
\end{figure*}

\subsubsection{ML-based Second Refinement Method} 
Based on the measurements in \eqref{secondlayerbt}, the ML estimation of channel parameters can be expressed as 
\begin{align}\label{ML}
	(\widehat{\Omega}_2,\widehat{b}_2,\widehat{g})\!=\!\arg\min_{\Omega,b,g} \sum_{m=-M_2}^{M_2}\!\big|\overline{y}_m^{*} \!-\!g\boldsymbol{\gamma}(\widetilde{\Theta}_m,\!\overline{k}_2)^{\rm H}\boldsymbol{\gamma}(\Omega,\!b)\big|^2
\end{align}
We  express \eqref{ML} in a vector form as
\begin{align}\label{ML2}
	(\widehat{\Omega}_2,\widehat{b}_2,\widehat{g})\!=\!\arg\min_{\Omega,b,g}\|\overline{\boldsymbol{y}} - g\boldsymbol{\Gamma}^{\rm H}\boldsymbol{\gamma}(\Omega,\!b)\|_2^2
\end{align}
where $\overline{\boldsymbol{y}}$ and $\boldsymbol{\Gamma}$ are the stack of $\overline{y}_m^*$  and $\boldsymbol{\gamma}(\widetilde{\Theta}_m,\!\overline{k}_2)$, respectively. Given $\boldsymbol{\gamma}(\Omega,\!b)$, the optimal solution of $g$ is $\widehat{g} = \boldsymbol{\gamma}(\Omega,\!b)^{\rm H}\boldsymbol{\Gamma}\overline{\boldsymbol{y}}/\|\boldsymbol{\Gamma}^{\rm H}\boldsymbol{\gamma}(\Omega,\!b)\|_2^2$. Then, \eqref{ML2} is converted to 
 \begin{align}\label{ML3}
 	(\widehat{\Omega}_2,\widehat{b}_2)\!=\!\arg\max_{\Omega,b} \big\|\overline{\boldsymbol{y}}^{\rm H}\boldsymbol{\Gamma}^{\rm H}\boldsymbol{\gamma}(\Omega,\!b)\big\|_2\big/\big\|\boldsymbol{\Gamma}^{\rm H}\boldsymbol{\gamma}(\Omega,\!b)\big\|_2
 \end{align}
which can be solved by the two-dimensional search. We term the THBT with this method as the THBT-ML.
 \subsubsection{PSP-based Second Refinement Method} 
From \eqref{coverageofk2_2}, the channel path locates in the beam coverage of $\boldsymbol{\gamma}(\widetilde{\Theta}_m,\!\overline{k}_2)$, which indicates that $|\boldsymbol{\gamma}(\Omega,\!b)^{\rm H}\boldsymbol{\gamma}(\widetilde{\Theta}_m,\!\overline{k}_2)|$ has similar values for $m=-M_2,\cdots,0,\cdots,M_2$. Therefore, we may normalize amplitudes of the measurements to simplify the analysis. Then we convert \eqref{ML} to 
\begin{align}\label{quasiML}
	(\widehat{\Omega}_2,\widehat{b}_2,\widehat{g})=\arg\min_{\Omega,b,g}\!\sum_{m=-M_2}^{M_2}\!(\mathcal{P}(\overline{y}_m^*) -\psi _m)^2
\end{align}
where $\psi _m \triangleq \mathcal{P}(g\boldsymbol{\gamma}(\widetilde{\Theta}_m,\!\overline{k}_2)^{\rm H}\boldsymbol{\gamma}(\Omega,\!b))$ and $\mathcal{P}(\cdot)$ denotes the phase of a complex number. One method to solve \eqref{quasiML} is performing the multi-dimensional search similar to \eqref{ML3}. However, a vast number of calculations are needed especially for intensive samples of $\Omega$ and $b$. To avoid the high computational complexity, we revisit the PSP in \eqref{PSPApp}. By extracting the phase of $G(\boldsymbol{u},\Omega,b)$, we have  
\begin{align}\label{PhasePSP}
	\mathcal{U}( G(\boldsymbol{u},\Omega,b))\!\approx\!\left\{ \begin{array}{ll}
		\!\frac{(\Omega-\Theta)^2}{4(k-b)}\pi\!-\!\frac{\pi}{4},&-\frac{N_{\rm t}}{2}\le z_0 \le \frac{N_{\rm t}}{2}\\
		0, &\!\mbox{others}
	\end{array} \right.
\end{align}
where $\mathcal{U}(\cdot)$ denotes the unwrapped phase of a complex number. By applying the PSP approximation, we can convert \eqref{quasiML} to  
\begin{align}\label{leastsquare}
	(\widehat{\Omega}_2,\!\widehat{b}_2,\!\widehat{\phi}_2)\!=\!\arg\min_{\Omega,b,\phi} \sum_{m=-M_2}^{M_2}\!\bigg(\!\mathcal{U}(\overline{y}_m^*)\!-\!\pi\frac{(\Omega\!-\!\widetilde{\Theta}_m)^2}{4(b\!-\!\overline{k}_2)} \!-\!\phi\!\bigg)^2
\end{align}
where  $\phi$ is a  parameter introduced to eliminate the constant phases. We define
\begin{align}
	\mathcal{L}(\alpha,\beta,\gamma) \triangleq \sum_{m=-M_2}^{M_2}\!(\mathcal{U}(\overline{y}_m^*)+ \alpha \widetilde{\Theta}_m^2 +\beta\widetilde{\Theta}_m +\gamma)^2
\end{align}
where $\alpha \triangleq -\frac{\pi}{4(b-\overline{k}_2)}$, $\beta \triangleq \frac{\pi\Omega}{2(b-\overline{k}_2)}$ and $\gamma \triangleq -\frac{\pi\Omega^2}{4(b-\overline{k}_2)}-\phi$. The optimal solution of \eqref{leastsquare} is achieved when
\begin{align}\label{Partial}
	&\frac{\partial\mathcal{L}(\alpha,\beta,\gamma) }{\partial (2\alpha)}\!= \!\sum_{m=-M_2}^{M_2}\!(\mathcal{U}(\overline{y}_m^*)\!+\!\alpha \widetilde{\Theta}_m^2 \!+\!\beta\widetilde{\Theta}_m \!+\!\gamma)\widetilde{\Theta}_m^2\!=\!0 \nonumber \\
	&\frac{\partial\mathcal{L}(\alpha,\beta,\gamma) }{\partial (2\beta)}\!= \!\sum_{m=-M_2}^{M_2}\!(\mathcal{U}(\overline{y}_m^*)\!+\!\alpha \widetilde{\Theta}_m^2 \!+\!\beta\widetilde{\Theta}_m \!+\!\gamma)\widetilde{\Theta}_m\!=\!0 \nonumber \\
	&\frac{\partial\mathcal{L}(\alpha,\beta,\gamma) }{\partial (2\gamma)}\!= \!\sum_{m=-M_2}^{M_2}\!(\mathcal{U}(\overline{y}_m^*)\!+\!\alpha \widetilde{\Theta}_m^2 \!+\!\beta\widetilde{\Theta}_m \!+\!\gamma)\!=\!0.
\end{align}
Note that \eqref{Partial} is a system of linear equations and the closed-form solutions can be obtained via the Gaussian elimination. We omit the details and denote its solutions as $\widehat{\alpha}$, $\widehat{\beta}$ and $\widehat{\gamma}$. Then we can express the results of the second refinement as
\begin{align}\label{Estimationofk}
\widehat{b}_2 = -\frac{\pi}{4\widehat{\alpha}} + \overline{k}_2,~\mbox{and}~\widehat{\Omega}_2 = -\frac{\widehat{\beta}}{2\widehat{\alpha}}.
\end{align}
We term the THBT with this method as the THBT-PSP.

\textbf{Remark 1:} One remaining problem is how to obtain the unwrapped phase of $\overline{y}_m^*$ in \eqref{leastsquare}. The classic method unwraps the phase by comparing the phase difference between the adjacent sample points with the threshold $\pi$~\cite{1982AOKI}. To guarantee the successful phase unwrap, we have 
\begin{align}\label{phasewrap}
	\frac{1}{\pi}|\mathcal{U}(\overline{y}_{m+1}^*) - \mathcal{U}(\overline{y}_{m}^*)| \le 1. 
\end{align}
Note that 
\begin{align}\label{phasewrap2}
	&~~~~\frac{1}{\pi}|\mathcal{U}(\overline{y}_{m+1}^*) - \mathcal{U}(\overline{y}_{m}^*)| \nonumber \\
	&\overset{\rm (a)}{\approx} \left|\frac{(\Omega\!-\!\widetilde{\Theta}_{m+1})^2}{4(b\!-\!\overline{k}_2)} - \frac{(\Omega\!-\!\widetilde{\Theta}_m)^2}{4(b\!-\!\overline{k}_2)}\right| \nonumber \\
	&=\frac{\overline{\Theta}_2|2\Omega - \widetilde{\Theta}_{m+1} - \widetilde{\Theta}_{m}|}{4|b\!-\!\overline{k}_2|}\nonumber \\
	&\overset{\rm (b)}{\le} \frac{\overline{\Theta}_2(1/N_{\rm t} + |b-\overline{k}_1|N_{\rm t} + M_2\overline{\Theta}_2)}{2|b\!-\!\overline{k}_2|}\nonumber \\
	&\overset{\rm (c)}{\le}\!\frac{\overline{\Theta}_2(1/N_{\rm t}\!+\!( \overline{b}-\widetilde{k}_1)N_{\rm t}\!+\!M_2\overline{\Theta}_2)}{2|b\!-\!\overline{k}_2|}
\end{align}
where we obtain $\rm (a)$ by applying the PSP approximation in \eqref{PhasePSP},  obtain $\rm (b)$ by considering the range of $\Omega$ in \eqref{initlalInterstedRegion21} and the expression of $\widetilde{\Theta}_m$ in \eqref{secondlayercodebookcodeword}, and obtain $\rm (c)$ by considering the range of $b$ in \eqref{initlalInterstedRegion21} and the expression of $\overline{k}_1$ in \eqref{widetildek2}.
Combining \eqref{phasewrap} and \eqref{phasewrap2}, we have 
\begin{align}\label{phasewrap3}
	\left\{ \begin{array}{ll}
	\overline{k}_2 \le  - B, &\overline{m}~\mbox{is even} \\
	\overline{k}_2 \ge  \overline{b} + B, &\overline{m}~\mbox{is odd}
\end{array} \right.
\end{align}
where 
\begin{align}
B\triangleq  \frac{\overline{\Theta}_2(1/N_{\rm t}\!+\!( \overline{b}-\widetilde{k}_1)N_{\rm t}\!+\!M_2\overline{\Theta}_2)}{2}.
\end{align}
With the setting of $\overline{k}_2$ in \eqref{phasewrap3}, the constraints in \eqref{phasewrap} are satisfied. Then, based on $\mathcal{P}(\overline{y}_m^*)$, the unwrapped phase of $\overline{y}_m^*$ can be expressed as
\begin{align}\label{phasewrap4}
\mathcal{U}(\overline{y}_{m+1}^*) =  \mathcal{U}(\overline{y}_{m}^*)\!+\!\mbox{mod}(\mathcal{P}(\overline{y}_{m+1}^*)\!-\!\mathcal{P}(\overline{y}_{m}^*)\!+\!\pi,2\pi)\!-\!\pi
\end{align}
for $m=-M_2,\cdots,0,\cdots,M_2-1$, where $\mathcal{U}(\overline{y}_{-M_2}^*)$ is initialized to be zero.

\textbf{Remark 2:} Note that two constraints on $\overline{k}_2$ are derived in \eqref{coverageofk2_2} and \eqref{phasewrap3}, respectively. To ensure the successful performing of the beam training, both \eqref{coverageofk2_2} and \eqref{phasewrap3} should be satisfied.  In addition, according to \eqref{absolutebeamgain}, a larger deviation between the surrogate distance of the channel and that of the codeword will result in a smaller beam gain. Therefore, we need to satisfy both \eqref{coverageofk2_2} and \eqref{phasewrap3} while reducing the deviation between the surrogate distance of the channel and that of the codeword. Based on the above discussions, we set  $\overline{k}_2$ as 
\begin{align}\label{k2setting}
	\overline{k}_2\!=\!\left\{ \begin{array}{ll}
		\!\min\{-B,\overline{k}_1 - M_2\overline{\Theta}_2/N_{\rm t} - 1/N_{\rm t}^2\},\!&\overline{m}~\mbox{is even} \\
		\!\max\{\overline{b} +B,\overline{k}_1 + M_2\overline{\Theta}_2/N_{\rm t}+ 1/N_{\rm t}^2\},\!&\overline{m}~\mbox{is odd}.
	\end{array} \right.
\end{align}

Now, we provide an example for the second-refinement codebook design and the beam training based on the PSP in \textbf{Design Example 2}. 

\textbf{Design Example 2:} Consider an XL-MIMO system with $N = 256$, $N_{\rm t} = 513$, $\lambda = 0.005$~m, $\overline{k}_1 = -6.09\times10^{-5}$, $\overline{b} = 1.22\times 10^{-4}$, $\overline{\Theta}_2 = 2/N_{\rm t}$, and $M_2 = 8$. The number of channel paths are set to $L =1$ and the corresponding channel steering vector is $\boldsymbol{\gamma}(\Omega,b)$, where $b=4.9782\times10^{-5}$ and $\Omega = 0.05$. Suppose the beam training result of the first refinement is $\overline{m} =0$. According to \eqref{initlalInterstedRegion2} and \eqref{initlalInterstedRegion21}, we can obtain potential region $\widetilde{\boldsymbol{\Phi}}_1$ and extended potential region $\boldsymbol{\Phi}_2$, which are illustrated as the red and the blue trapezoidal regions in Fig.~\ref{Coverage}, respectively. According to \eqref{k2setting}, we can obtain $\overline{k}_2$. Then, we can design $\boldsymbol{w}_m$ based on \eqref{secondlayercodebookcodeword}.  As shown in Fig.~\ref{Coverage}, the beam coverage of $\boldsymbol{w}_{m+1}$ can be obtained through translating the beam coverage of $\boldsymbol{w}_{m}$ by $\overline{\Theta}_2$.  In Fig.~\ref{PhaseUnwarp},  we illustrate the wrapped phase, the unwrapped phase, and the approximated phase of $\overline{y}_m^*$ via the PSP. It is shown that the phase of $\overline{y}_m^*$ is well unwrapped via \eqref{phasewrap4} and the unwrapped phase can be well approximated by the PSP. In Fig.~\ref{ParametersEstimation}, we illustrate the beam gain of $\boldsymbol{\gamma}(\Omega,b)$ and the values  of the real parameters as well as the estimated parameters in \eqref{Estimationofk}. It is shown that the estimated parameters are close to the real parameters, which verifies the effectiveness of the second refinement.

Finally, we summarize the ML-based and PSP-based second refinement methods in \textbf{Algorithm~\ref{Alg2}}.

\begin{algorithm}[!t]
	\caption{ML-based and PSP-based Second Refinement Method}
	\label{Alg2}
	\begin{algorithmic}[1]
		\STATE \textbf{Input:} $N$, $N_{\rm t}$, $\lambda$, $\boldsymbol{h}$, $\overline{m}$, $\overline{b}$, $\overline{k}_1$, $\widetilde{k}_1$, $M_2$, $\overline{\Theta}_2$.
		\STATE Obtain $\overline{k}_2$ via \eqref{k2setting}.
		\STATE Obtain $\boldsymbol{\mathcal{C}}_2$ via \eqref{secondlayercodebook}.
		\STATE Obtain $\overline{y}_m$ for $m=-M_2,\cdots,0,\cdots,M_2$ via \eqref{secondlayerbt}.
		\IF{ML is adopted}   
		\STATE Obtain $\widehat{\Omega}_2$ and $\widehat{b}_2$ via \eqref{ML3}.
		\ELSIF{PSP is adopted}
		\STATE Obtain $\mathcal{U}(\overline{y}_{m}^*)$ via \eqref{phasewrap4}.
		\STATE Obtain $\widehat{\Omega}_2$ and $\widehat{b}_2$ via \eqref{Estimationofk}.
		\ENDIF
		\STATE \textbf{Output:} $\widehat{\Omega}_2$ and $\widehat{b}_2$.
	\end{algorithmic}
\end{algorithm}

%

\section{GA-based Third Refinement Method}\label{SecThirdLayer}
In this section, we develop the GA-based third refinement method to further improve the beam training accuracy. First, we perform the hybrid-field neighboring search to identify the potential region of the main lobe of the channel steering vector. Then, by applying the GA, an LS estimator is developed to obtain the high-accuracy channel parameter estimation.

\subsection{Hybrid-Field Neighboring Search}\label{SecNS}

In this part, we present the details of the hybrid-field neighboring search, which includes initialization, beam training, and stop conditions.

\subsubsection{Initialization} We divide the hybrid-field neighboring search into $M_{\rm n}$ groups, each consisting of five times of beam training. The parameters of the central codeword in the neighboring search are initialized to be $\widetilde{\Theta}_{\rm n}^{(1)} = \widehat{\Omega}_2$  and $\widetilde{k}_{\rm n}^{(1)} = \widehat{b}_2$. The surrogate distance deviation and the angle deviation between adjacent codewords are set to $\overline{k}_{\rm n}$ and $\overline{\Theta}_{\rm n}$, respectively. The index  of  the current group is initialized to be $m=0$. 

\subsubsection{Beam Training}\label{ThirdLyaersubBeamtraining}
 We update $m\leftarrow m+1$ and design the codebook for the beam training of the $m$th group as 
\begin{align}\label{thirdlayersubcodebook}
	\boldsymbol{\mathcal{C}}_{\rm n}^{(m)}\!=\!\big\{	&\boldsymbol{\gamma}\big(\widetilde{\Theta}_{\rm n}^{(m)}\!-\!\overline{\Theta}_{\rm n},\widetilde{k}_{\rm n}^{(m)}\big),
	\boldsymbol{\gamma}\big(\widetilde{\Theta}_{\rm n}^{(m)}\!+\!\overline{\Theta}_{\rm n},\widetilde{k}_{\rm n}^{(m)}\big),\nonumber \\
	&\boldsymbol{\gamma}\big(\widetilde{\Theta}_{\rm n}^{(m)},\widetilde{k}_{\rm n}^{(m)}\!-\!\overline{k}_{\rm n}\big),
	\boldsymbol{\gamma}\big(\widetilde{\Theta}_{\rm n}^{(m)},\widetilde{k}_{\rm n}^{(m)}\!+\!\overline{k}_{\rm n}\big),\nonumber \\
	&~~~~~~~~~~~~~~~~~~~~~~~~~~~~~~\boldsymbol{\gamma}\big(\widetilde{\Theta}_{\rm n}^{(m)},\widetilde{k}_{\rm n}^{(m)}\big)\big\}.
\end{align}
After that, we perform beam training with $\boldsymbol{\mathcal{C}}_{\rm n}^{(m)}$ and obtain 
\begin{align}\label{thirdlayerbt}
	\widetilde{y}_s^{(m)} = \boldsymbol{h}^{\rm H}\{\boldsymbol{\mathcal{C}}_{\rm n}^{(m)}\}_s + \eta
\end{align}
for $s=1,2,3,4,5$. The index of the codeword in $\boldsymbol{\mathcal{C}}_{\rm n}^{(m)}$ with the biggest received power can be expressed as 
\begin{align}\label{thirdlayerbtre}
	\widetilde{s}^{(m)} = \arg\max_{s = 1,2,3,4,5} |\widetilde{y}_s^{(m)} |.
\end{align}
Then we update the parameters of the central codeword as 
\begin{align}\label{centralparameter}
	\big(\widetilde{\Theta}_{\rm n}^{(m+1)},\widetilde{k}_{\rm n}^{(m+1)}\big) = \varXi(\{\boldsymbol{\mathcal{C}}_{\rm n}^{(m)}\}_{\widetilde{s}^{(m)}})
\end{align}
where $\varXi(\cdot)$ denotes the channel parameters of a steering vector, e.g., $(\Omega,b) = \varXi(\boldsymbol{\gamma}(\Omega,b))$.
\subsubsection{Stop Conditions}\label{ThirdLyaerStopConditions}
 We repeat the beam training procedure in Section \ref{ThirdLyaersubBeamtraining} until $\widetilde{s}^{(m)} =5$ or $m = M_{\rm n}$, where the index of the training group at this point is expressed as $\widetilde{m}$. Note that $\widetilde{s}^{(\widetilde{m})} = 5$ implies that the neighboring search has converged due to $\widetilde{\Theta}_{\rm n}^{(\widetilde{m}+1)} = \widetilde{\Theta}_{\rm n}^{(\widetilde{m}+2)}$ and $\widetilde{k}_{\rm n}^{(\widetilde{m}+1)} = \widetilde{k}_{\rm n}^{(\widetilde{m}+2)}$. We express the neighboring search results as
 \begin{align}\label{thirdlayerbtre2}
 	\widehat{\Omega}_{\rm n} = \widetilde{\Theta}_{\rm n}^{(\widetilde{m}+1)},~\mbox{and}~\widehat{b}_{\rm n} = \widetilde{k}_{\rm n}^{(\widetilde{m}+1)}.
 \end{align}


\textbf{Remark 3:} Usually, $\overline{k}_{\rm n}$ and $\overline{\Theta}_{\rm n}$ can be designed according to the coherence of the adjacent codewords, which is equivalent to the main lobe region determination of the channel steering vector.  Note that the coherence of the two codewords that are adjacent in angle equals that of the two  adjacent discrete Fourier transform (DFT) codewords~\cite{Tcom22CMH}. Therefore, we can set $\overline{\Theta}_{\rm n} = 2/N_{\rm t}$ emulating the merits of the DFT codebook~\cite{TWC20QCH}. The coherence of the two codewords that are adjacent in distance can be calculated as 
\begin{align}\label{DistanceCoherence}
	\rho &= \frac{1}{N_{\rm t}} |G(\boldsymbol{u},\Theta,k+\overline{k}_{\rm n})|\nonumber \\
	& \overset{\rm (a)}{\approx}  \frac{1}{N_{\rm t}}\left|\int_{-N}^{N}e^{j\pi \overline{k}_{\rm n}z^2}dz \right|\nonumber \\
	& = \frac{1}{N_{\rm t}}\sqrt{\frac{2\mathcal{C}(\sqrt{2\overline{k}_{\rm n}}N)^2 + 2\mathcal{S}(\sqrt{2\overline{k}_{\rm n}}N)^2}{\overline{k}_{\rm n}}} 
\end{align}
where we approximate the summation as integral in $\rm (a)$, $\mathcal{C}(\sqrt{2\overline{k}_{\rm n}}N) = \int_0^{\sqrt{2\overline{k}_{\rm n}}N} \cos(\pi z^2/2){\rm d} z$ and $\mathcal{S}(\sqrt{2\overline{k}_{\rm n}}N) = \int_0^{\sqrt{2\overline{k}_{\rm n}}N} \sin(\pi z^2/2){\rm d} z$ are the Fresnel functions. Given $\rho$, the value of $\overline{k}_{\rm n}$ can be obtained via \eqref{DistanceCoherence}. For example, if $N_{\rm t} = 513$ and $\rho = 0.35$, we can obtain $\overline{k}_{\rm n} = 2.28\times 10^{-5}\approx 6/N_{\rm t}^2$.
\begin{figure}[!t]
	\centering
	\includegraphics[width=83.2mm]{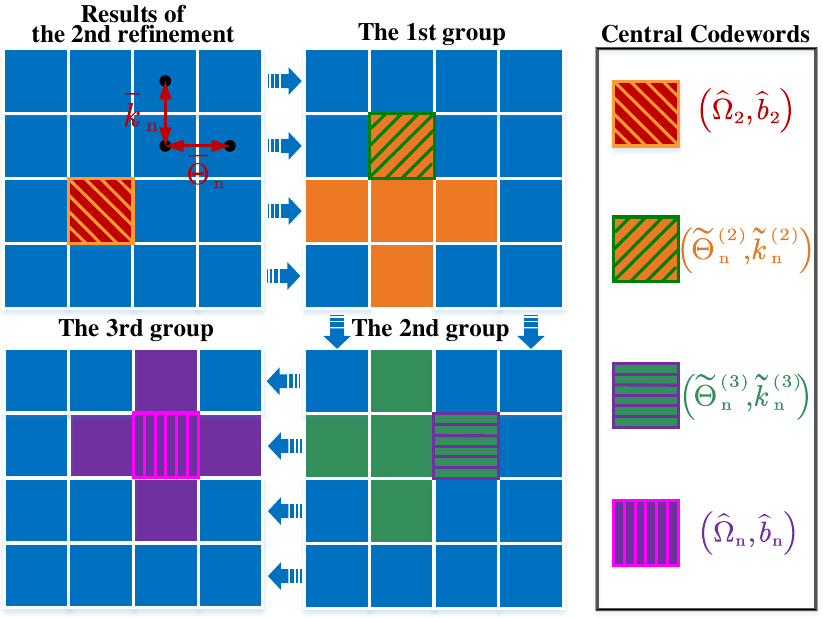}
	\caption{Illustration of the hybrid-field neighboring search.}
	\label{ThirdLayerExample}
\end{figure}

\textbf{Remark 4:} From Section \ref{ThirdLyaerStopConditions}, the procedure of the hybrid-field neighboring search stops when $\widetilde{s}^{(\widetilde{m})} =5$ or $\widetilde{m} = M_{\rm n}$. If $\widetilde{s}^{(\widetilde{m})} =5$ ($\widetilde{m}$ equals $M_{\rm n}$ or not), which implies that the central codeword has the largest received power, we say that the neighboring search is successful. Then we can implement the GA to further improve the training performance.  Otherwise, we say that the neighboring search fails. Then the THBT stops at this step and its results are expressed as 
\begin{align}
\widehat{b}_{\rm f} =\widehat{b}_{\rm n},~\mbox{and}~\widehat{\Omega}_{\rm f} = \widehat{\Omega}_{\rm n}.
\end{align}

Now, we provide an example for the hybrid-field neighboring search in \textbf{Design Example 3}.

\textbf{Design Example 3:} In Fig.~\ref{ThirdLayerExample}, we consider the neighboring search consisting of $M_{\rm n} =3$ groups, where each group tests the adjacent four codewords of the central codeword. If the best codeword in the third group is the one with pink borders, the neighboring search is successful. Otherwise, it fails. In addition, it is noteworthy that there are two overlapping  codewords between the adjacent groups. Therefore, each group only needs to test three additional codewords except the first one.  In total, the maximum training overhead of the hybrid-field neighboring search is $3M_{\rm n} +2$.

\begin{figure}[!t]
	\centering
	\includegraphics[width=83.5mm]{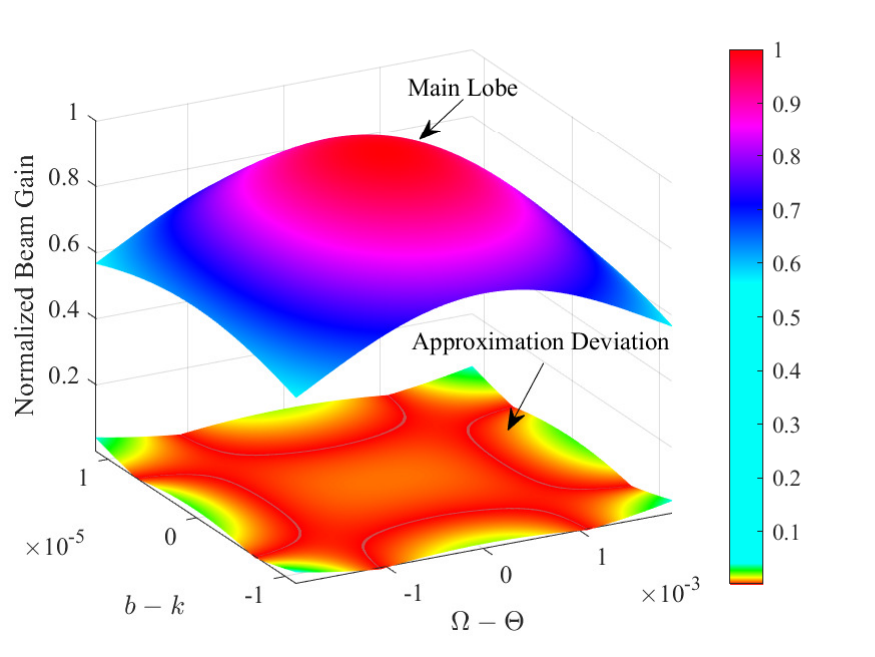}
	\caption{Illustration of the Gaussian approximation.}
	\label{GaussianApproximation}
\end{figure}

\subsection{Gaussian Approximation}\label{SecGauAppro}
From \eqref{BeamGain}, we have 
\begin{align}\label{AbsoluteBeamGain}
	|G(\boldsymbol{u},\Omega,b)|^2 &= \left(\sum_{n=-N}^{N}e^{j( \widetilde{b}n^2-\widetilde{\Omega}n )}\right)\!\left(\sum_{n=-N}^{N}e^{j( \widetilde{b}n^2-\widetilde{\Omega}n )}\right)^* \nonumber \\
	&=\sum_{n=-N}^{N}\sum_{i=-N}^{N}e^{j\big(\widetilde{\Omega}(i-n) + \widetilde{b}(n^2-i^2)\big)}\nonumber \\
	&=\sum_{n=-N}^{N}\sum_{i=-N}^{N}\cos\big(\widetilde{\Omega}(i-n) + \widetilde{b}(n^2-i^2)\big)
\end{align} 
where $\widetilde{\Omega} \triangleq  (\Omega-\Theta)\pi$ and $\widetilde{b} \triangleq (b-k)\pi$. In fact, it is hard to obtain a closed-form expression of \eqref{AbsoluteBeamGain} due to the quadratic phase term $\widetilde{b}(n^2-i^2)$. To obtain a deeper scope of $|G(\boldsymbol{u},\Omega,b)|^2$, we resort to the widely-used Taylor series. 
\begin{figure*}[htbp]
	\begin{align}\label{TaylorSeries1}
		|G(\boldsymbol{u},\Omega,b)|^2\! &=\!\sum_{n=-N}^{N}\sum_{i=-N}^{N}\sum_{p=0}^{\infty}\sum_{q=1}^{2p-1}\! \frac{(-1)^p(i\!-\!n)^{2p}\widetilde{\Omega}^{2p}}{(2p)!}\!+\! \frac{(-1)^p(n^2\!-\!i^2)^{2p}\widetilde{b}^{2q}}{(2p)!}\! +\!2\frac{(-1)^{p}(i-n)^{q}(n^2\!-\!i^2)^{2p-q}\widetilde{\Omega}^{q}\widetilde{b}^{2p-q}}{(q)!(2p-q)!} \nonumber \\
		&\overset{(\rm a)}{\approx}\sum_{p=0}^{\infty}\sum_{q=1}^{2p-1}\! \frac{(-1)^pC_{p}N^{2p}\widetilde{\Omega}^{2p}}{(2p)!}\!+\! \frac{(-1)^p\widetilde{C}_{p}(N^2)^{2p}\widetilde{b}^{2q}}{(2p)!}\! +\!2\frac{(-1)^{p}\overline{C}_{p,q}N^{4p-q+2}\widetilde{\Omega}^{q}\widetilde{b}^{2p-q}}{(q)!(2p-q)!}
	\end{align}
	\hrulefill
\end{figure*}
\begin{figure*}[htbp]
	\begin{align}\label{TaylorSeries2}
		e^{\mathlarger{-\frac{\widetilde{\Omega}^2}{\sigma_1^2}-\frac{\widetilde{b}^2}{\sigma_2^2}}}\! &=\sum_{p=0}^{\infty}\sum_{q=1}^{p-1}\! \frac{(-1)^p\big(\frac{2}{\sigma_1^2}\big)^p\widetilde{\Omega}^{2p}\prod_{i=1}^{p}(2i-1)}{(2p)!}+ \frac{(-1)^p\big(\frac{2}{\sigma_2^2}\big)^p\widetilde{b}^{2q}\prod_{i=1}^{p}(2i-1)}{(2p)!}\nonumber \\ &~~~~~~~~~~~~+2\frac{(-1)^{p}\big(\frac{2}{\sigma_1^2}\big)^{q}\big(\frac{2}{\sigma_2^2}\big)^{p-q}\big(\prod_{i=1}^{q}(2i-1)\big)\big(\prod_{i=1}^{p-q}(2i-1)\big) \widetilde{\Omega}^{2q}\widetilde{b}^{2p-2q}}{(2q)!(2p-2q)!} \nonumber \\
		&=\sum_{p=0}^{\infty}\sum_{q=1}^{p-1}\! \frac{(-1)^pD_{p}\left(\sigma_1^{-1}\right)^{2p}\widetilde{\Omega}^{2p}}{(2p)!}\!+\! \frac{(-1)^p\widetilde{D}_{p}\left(\sigma_2^{-1}\right)^{2p}\widetilde{b}^{2q}}{(2p)!}\! +\!2\frac{(-1)^{p}\overline{D}_{p,q}\left(\sigma_1^{-1}\right)^{2q}\left(\sigma_2^{-1}\right)^{2p-2q}\widetilde{\Omega}^{2q}\widetilde{b}^{2p-2q}}{(2q)!(2p-2q)!}
	\end{align}
	\hrulefill
	\vspace*{-0.1cm}
\end{figure*}
At the top of the next page, we provide the Taylor series of $|G(\boldsymbol{u},\Omega,b)|^2$ and $e^{-{\widetilde{\Omega}^2}/{\sigma_1^2}-{\widetilde{b}^2}/{\sigma_2^2}}$ about the point $(0,0)$, where $C_p$,~$\widetilde{C}_p$,  $\overline{C}_{p,q}$, $D_p$,~$\widetilde{D}_p$,  $\overline{D}_{p,q}$ are all constant coefficients. In addition, we omit lower-order terms in $(\rm a)$ of \eqref{TaylorSeries1} for simplicity. From \eqref{TaylorSeries1} and \eqref{TaylorSeries2}, the Taylor series of $|G(\boldsymbol{u},\Omega,b)|^2$ and $e^{-{\widetilde{\Omega}^2}/{\sigma_1^2}-{\widetilde{b}^2}/{\sigma_2^2}}$ share similarities in several aspects including the orders and coefficients of the series. Inspired by their similarities, we use the two-dimensional Gaussian function
\begin{align}\label{GaussianFunction}
	f(\Omega,b) \triangleq ae^{\mathlarger{-\frac{(\Omega-\Theta)^2}{2\sigma_1^2}-\frac{(b-k)^2}{2\sigma_2^2}}}
\end{align}
to approximate the main lobe of $|G(\boldsymbol{u},\Omega,b)|$, which is termed as the Gaussian approximation. Note that the Gaussian approximation has been investigated in~\cite{TVT23ZA,TAP17BG,TAES99LLW}. However, they only focus on the far-field scenarios. In this work, we demonstrate that the main lobe of the HFBG can also be approximated by the Gaussian functions. According to the \textbf{Remark 3} of Section~\ref{SecThirdLayer}, the main lobe of $|G(\boldsymbol{u},\Omega,b)|$ is restricted to 
\begin{align}
	\left\{(\Omega,b)\big| \Omega\!\in\!\left[\Theta\!-\!\overline{\Theta}_3,\Theta\!+\!\overline{\Theta}_3\right],~b\!\in\!\left[k\!-\!\overline{k}_3,k\!+\!\overline{k}_3\right]\right\}
\end{align}
where $\overline{\Theta}_3 = \overline{\Theta}_{\rm n}/2$ and $\overline{k}_3 = \overline{k}_{\rm n}/2$.  Then, the Gaussian approximation is formulated as
\begin{align}\label{GaussianApprox}
	&\min_{a,\sigma_1,\sigma_2}\!\int_{k-\overline{k}_3}^{k+\overline{k}_3}\int_{\Theta-\overline{\Theta}_3}^{\Theta+\overline{\Theta}_3}\big|f(\Omega,b)\!-\!|G(\boldsymbol{u},\!\Omega,\!b)|\big|^2 {\rm d}\Omega{\rm d}b 
\end{align}
which is a nonlinear least-square problem and can be solved by the trust-region optimization algorithm~\cite{SJO96TF}. We omit the details and denote the solutions of \eqref{GaussianApprox} as $\widehat{a}$, $\widehat{\sigma}_1$ and $\widehat{\sigma}_2$. Then, the optimized Gaussian function can be expressed as 
\begin{align}\label{GaussianFunction2}
	\widehat{f}(\Omega,b) = \widehat{a}e^{\mathlarger{-\frac{(\Omega-\Theta)^2}{2\widehat{\sigma}_1^2}-\frac{(b-k)^2}{2\widehat{\sigma}_2^2}}}.
\end{align}

Note that for another channel steering vector, $\overrightarrow{\boldsymbol{u}} = \boldsymbol{\gamma}(\overrightarrow{\Theta},\overrightarrow{k})$, we have  
\begin{align}\label{BeamGain2}
	G(\overrightarrow{\boldsymbol{u}},\Omega,b) &= \sum_{n=-N}^{N}e^{j\pi((\overrightarrow{\Theta}-\Omega)n - (\overrightarrow{k}-b)n^2)} \nonumber \\
	&= \sum_{n=-N}^{N}e^{j\pi((\Theta-(\Theta-\overrightarrow{\Theta}+\Omega))n - (k-(k-\overrightarrow{k}+b))n^2)} \nonumber \\
	&= G(\boldsymbol{u},\Omega + (\Theta-\overrightarrow{\Theta}),b + (k-\overrightarrow{k}))
\end{align}
which indicates that the beam gain of $\overrightarrow{\boldsymbol{u}}$ is the translation of that of $\boldsymbol{u}$. Therefore, we only need to solve $\eqref{GaussianApprox}$ for $\boldsymbol{u}$, and the Gaussian approximation for other channel steering vectors can be obtained via the translation in \eqref{BeamGain2}.

In Fig.~\ref{GaussianApproximation}, we illustrate the main lobe of $|G(\boldsymbol{u},\Omega,b)|$ and the deviation of the Gaussian approximation, where we set $N=256$,~$N_{\rm t}=513$, $\overline{k}_3 = 3/N_{\rm t}^2$, and $\overline{\Theta}_3 = 1/N_{\rm t}$. From the figure, the approximation deviation is quite small compared to the beam gain of the main lobe. For example, the maximum and the averaged approximation deviations are only $4\%$  and $0.5\%$ of the maximum beam gain, respectively.

\subsection{GA-based Channel Parameter Estimation}
In this part, we implement the channel parameter estimation based on Gaussian approximation to improve the estimation accuracy with low overhead.

First, we determine the potential region of $\Omega$ and $b$ based on the results of the neighboring search. From \eqref{thirdlayerbtre2}, the potential intervals of $\Omega$ and $b$ can be expressed as
\begin{align}
	&\boldsymbol{\Omega} = \left[\widehat{\Omega}_{\rm n} -\overline{\Theta}_3, \widehat{\Omega}_{\rm n} + \overline{\Theta}_3\right],\nonumber \\
	&\boldsymbol{\mathcal{B}} = \left[\widehat{b}_{\rm n} -\overline{k}_3,  \widehat{b}_{\rm n} + \overline{k}_3\right].
\end{align}

Then, we further narrow down the regions of $\Omega$ and $b$ by comparing the powers of the received signals adjacent to $\widetilde{y}_5^{(\widetilde{m})}$. The narrowed potential region of $\Omega$ is expressed as 
\begin{align}\label{EqTheta}
	\widetilde{\boldsymbol{\Omega}} = \left\{ \begin{array}{ll}
		\left[\widehat{\Omega}_{\rm n},\widehat{\Omega}_{\rm n} + \overline{\Theta}_3\right],\!&\!|\widetilde{y}_2^{(\widetilde{m})}|\geq|\widetilde{y}_1^{(\widetilde{m})}|\vspace{1ex} \\
		\left[\widehat{\Omega}_{\rm n} - \overline{\Theta}_3,\widehat{\Omega}_{\rm n}\right], &|\widetilde{y}_2^{(\widetilde{m})}|<|\widetilde{y}_1^{(\widetilde{m})}|
	\end{array} \right.
\end{align}
where the conditions can be obtained conveniently according to \eqref{thirdlayersubcodebook} and \eqref{thirdlayerbt}. Similarly, the narrowed  potential region of $b$ can be expressed as
\begin{align}\label{EqK}
	\widetilde{\boldsymbol{\mathcal{B}}} = \left\{ \begin{array}{ll}
		\left[\widehat{b}_{\rm n},  \widehat{b}_{\rm n} + \overline{k}_3\right],\!&\!|\widetilde{y}_4^{(\widetilde{m})}|\geq|\widetilde{y}_3^{(\widetilde{m})}|\vspace{1ex}\\
		\left[\widehat{b}_{\rm n} -\overline{k}_3,  \widehat{b}_{\rm n}\right], &|\widetilde{y}_4^{(\widetilde{m})}|<|\widetilde{y}_3^{(\widetilde{m})}|.
	\end{array} \right.
\end{align}

After that, we quantize the narrowed  potential regions of $\Omega$ and $b$ by $M_3$ samples. We denote the left and right boundaries of $\widetilde{\boldsymbol{\Omega}}$ as $\Psi_{\rm L}$ and $\Psi_{\rm R}$, respectively. Similarly, we have $\widetilde{\boldsymbol{\mathcal{B}}} = [d_{\rm L},d_{\rm R}]$. Then the $m$th sample of the angle and the surrogate distance can be expressed as 
\begin{align}\label{quantizedtk}
	&\widetilde{\Theta}_3^{(m)} =  \Psi_{\rm L} + \frac{(m-1)(\Psi_{\rm R}-\Psi_{\rm L})}{M_3-1}, \nonumber \\
	&\widetilde{k}_3^{(m)} =  d_{\rm L} + \frac{(m-1)(d_{\rm R}-d_{\rm L})}{M_3-1}.
\end{align}
We stack the codewords directing to the quantized samples in \eqref{quantizedtk} as $\boldsymbol{\mathcal{C}}_3$, where 
\begin{align}\label{fourthlayercodebook}
	&\boldsymbol{\mathcal{C}}_3\!=\! \left\{\widetilde{\boldsymbol{\mathcal{C}}}_3^{(1)},\widetilde{\boldsymbol{\mathcal{C}}}_3^{(2)},\cdots,\widetilde{\boldsymbol{\mathcal{C}}}_3^{(M_3)}\right\},\nonumber \\
	 &\big[\widetilde{\boldsymbol{\mathcal{C}}}_3^{(m)}\big]_t\!=\!\boldsymbol{\gamma}\big(\widetilde{\Theta}_3^{(m)},\widetilde{k}_3^{(t)}\big)
\end{align}
for $m = 1,2,\cdots,M_3$ and $t=1,2,\cdots,M_3$. Then we perform the beam training with $\boldsymbol{\mathcal{C}}_3$. Similar to \eqref{secondlayerbt}, the received signal can be expressed as 
\begin{align}\label{ReceivedSignal}
	\widehat{y}_{m,t}& \approx g^*\boldsymbol{\gamma}(\Omega,b)^{\rm H}\boldsymbol{\gamma}\big(\widetilde{\Theta}_3^{(m)},\widetilde{k}_3^{(t)}\big) + \eta.
\end{align} 
We specify that all the variants of $\eta$ denote the noise terms in the following texts. 
\begin{algorithm}[!t]
	\caption{GA-based Third Refinement Method}
	\label{alg3}
	\begin{algorithmic}[1]
		\STATE \textbf{Input:} $N$, $N_{\rm t}$, $\lambda$, $\boldsymbol{h}$, $\widehat{\Omega}_2$, $\widehat{b}_2$, $M_{\rm n}$, $M_3$.		
     	\STATE Obtain $\overline{k}_{\rm n}$ and $\overline{\Theta}_{\rm n}$ via \textbf{Remark 3}.
     	\STATE Obtain $\widetilde{m}$ and $\widetilde{s}^{(\widetilde{m})}$ via Section \ref{ThirdLyaerStopConditions}.
     	\STATE Obtain $\widehat{b}_{\rm n}$ and $\widehat{\Omega}_{\rm n}$ via \eqref{thirdlayerbtre2}.
		\IF{$\widetilde{m} = M_{\rm n}$ and $\widetilde{s}^{(\widetilde{m})}\neq5$}
		\STATE $\widehat{b}_{\rm f} \leftarrow \widehat{b}_{\rm n}$, $\widehat{\Omega}_{\rm f} \leftarrow \widehat{\Omega}_{\rm n}$.
		\ELSE
		\STATE Obtain $\boldsymbol{\mathcal{C}}_3$ via \eqref{fourthlayercodebook}.
		\STATE Obtain $\widehat{y}_{m,t}$ via \eqref{ReceivedSignal}.
		\STATE Obtain $\widehat{b}_{3}$ and $\widehat{\Omega}_{3}$ via \eqref{LSSolution}.
		\STATE $\widehat{b}_{\rm f} \leftarrow \widehat{b}_3$, $\widehat{\Omega}_{\rm f} \leftarrow \widehat{\Omega}_3$.
		\ENDIF
		\STATE Obtain $\widehat{r}_{\rm f}$ via \eqref{DistanceEstimation}.
		\STATE \textbf{Output:}  $\widehat{b}_{\rm f}$, $\widehat{\Omega}_{\rm f}$, $\widehat{r}_{\rm f}$.
	\end{algorithmic}
\end{algorithm}

By applying the Gaussian approximation to \eqref{ReceivedSignal}, we have 
\begin{align}\label{ReceivedSignal2}
	|\widehat{y}_{m,t}| &\approx |\widetilde{g}|\widehat{f}(\widetilde{\Theta}_3^{(m)},\widetilde{k}_3^{(t)}) + \widetilde{\eta} \nonumber \\ 
	&=|\widetilde{g}|\widehat{a}e^{\mathlarger{-\frac{(\Omega-\widetilde{\Theta}_3^{(m)})^2}{2\widehat{\sigma}_1^2}-\frac{(b-\widetilde{k}_3^{(t)})^2}{2\widehat{\sigma}_2^2}}} + \widetilde{\eta}
\end{align} 
where $\widetilde{g} = g^*/N_{\rm t}$. Note that the Gaussian function is the exponential of the quadratic function. A straightforward method to simplify the analysis is to take  the natural logarithm of $|\widehat{y}_{m,t}|$. Then, we have 
\begin{align}\label{ReceivedSignal3}
	\ln |\widehat{y}_{m,t}| = \ln(|\widetilde{g}|\widehat{a}) \!-\!\frac{(\Omega-\widetilde{\Theta}_3^{(m)})^2}{2\widehat{\sigma}_1^2}-\frac{(b-\widetilde{k}_3^{(t)})^2}{2\widehat{\sigma}_2^2}\!+\!\overline{\eta},
\end{align} 
which is  a quadratic function about the channel parameters $\Omega$ and $b$. Note that
\begin{align}\label{OneOrderTaylor}
	\ln |\widehat{y}_{m,t}| &\approx \ln(|\widetilde{g}|\widehat{f}(\widetilde{\Theta}_3^{(m)},\widetilde{k}_3^{(t)})) + \ln\bigg(1+\frac{\widetilde{\eta}}{|\widetilde{g}|\widehat{f}(\widetilde{\Theta}_3^{(m)},\widetilde{k}_3^{(t)})}\bigg)\nonumber \\
	&\overset{\rm (a)}{\approx} \ln(|\widetilde{g}|\widehat{f}(\widetilde{\Theta}_3^{(m)},\widetilde{k}_3^{(t)})) + \frac{\widetilde{\eta}}{|\widetilde{g}|\widehat{f}(\widetilde{\Theta}_3^{(m)},\widetilde{k}_3^{(t)})}
\end{align}
where $\rm (a)$ holds because $\ln(1+\epsilon)\approx\epsilon$. The relations in \eqref{OneOrderTaylor} indicate that the noise term will be magnified by $\frac{1}{|\widetilde{g}|\widehat{f}(\widetilde{\Theta}_3^{(m)},\widetilde{k}_3^{(t)})}$ times in the process of taking the logarithm. Therefore, large errors will be introduced for small values of $\widehat{f}(\widetilde{\Theta}_3^{(m)},\widetilde{k}_3^{(t)})$. To avoid the noise amplification effects, we multiply \eqref{ReceivedSignal3} by $|\widehat{y}_{m,t}|$~\cite{SPM11_GHW,SPM22_WK}, and obtain
\begin{align}\label{WlSRaw}
	\varUpsilon_{m,t} =  \frac{|\widehat{y}_{m,t}|\widetilde{\Theta}_3^{(m)}}{\widehat{\sigma}_1^2}\Omega  + \frac{|\widehat{y}_{m,t}|\widetilde{k}_3^{(t)}}{\widehat{\sigma}_2^2}b  +|\widehat{y}_{m,t}|\chi + \widehat{\eta}
\end{align}
where 
\begin{align}\label{WlSRawVar}
	&\varUpsilon_{m,t} =  |\widehat{y}_{m,t}|\ln |\widehat{y}_{m,t}| + \frac{|\widehat{y}_{m,t}|(\widetilde{\Theta}_3^{(m)})^2}{2\widehat{\sigma}_1^2} + \frac{|\widehat{y}_{m,t}|(\widetilde{k}_3^{(t)})^2}{2\widehat{\sigma}_2^2} \nonumber \\
	&\chi = \ln(|\widetilde{g}|\widehat{a}) - \frac{\Omega^2}{2\widehat{\sigma}_1^2} - \frac{b^2}{2\widehat{\sigma}_2^2}.
\end{align}
We rewrite \eqref{WlSRaw} in the vector form as 
\begin{align}\label{WLSproblem}
	\boldsymbol{A} \boldsymbol{z} + \widehat{\boldsymbol{\eta}} = \boldsymbol{y},
\end{align}
where $\boldsymbol{z}\triangleq[\Omega,b,\chi]^{\rm T}$ and $\widehat{\boldsymbol{\eta}}$ is the stack of noise terms. The $u$th row of $\boldsymbol{A}$ and the $u$th entry of $\boldsymbol{y}$ are expressed as 
\begin{align}
	&[\boldsymbol{A}]_{u,:}\!=\! \bigg[\frac{|\widehat{y}_{m,t}|\widetilde{\Theta}_3^{(m)}}{\widehat{\sigma}_1^2},\frac{|\widehat{y}_{m,t}|\widetilde{k}_3^{(t)}}{\widehat{\sigma}_2^2},|\widehat{y}_{m,t}|\bigg],~[\boldsymbol{y}]_u\!=\!\varUpsilon_{m,t},
\end{align}
respectively, for $u = (m-1)M_3 +t$, $m=1,2,\cdots,M_3$, and $t=1,2,\cdots,M_3$. The LS solution of \eqref{WLSproblem} is $\widehat{\boldsymbol{z}} = (\boldsymbol{A}^{\rm T} \boldsymbol{A})^{-1}\boldsymbol{A}^{\rm T}\boldsymbol{y}$. Then the estimates of $\Omega$ and $b$ can be expressed as 
\begin{align}\label{LSSolution}
\widehat{\Omega}_{3} = [\widehat{\boldsymbol{z}}]_1,~\mbox{and}~\widehat{b}_{3} = [\widehat{\boldsymbol{z}}]_2.
\end{align}

\begin{figure}[!t]
	\centering
	\includegraphics[width=85mm]{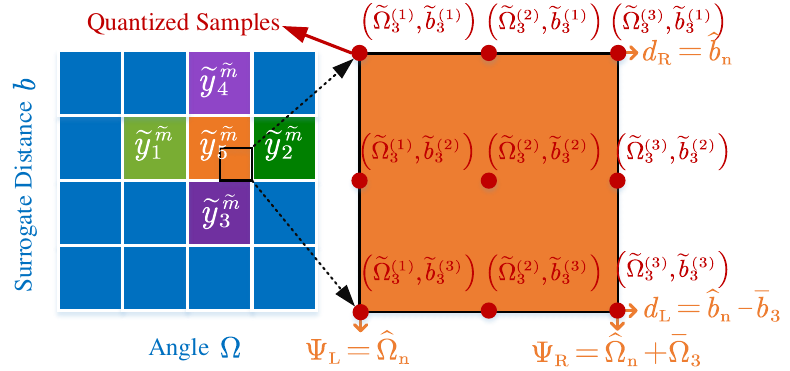}
	\caption{Illustration of the GA-based channel parameter estimation.}
	\label{FourthLayerExample}
\end{figure}

%

Now we provide an example for the GA-based channel parameter estimation in \textbf{Design Example 4}.

\textbf{Design Example 4:} In Fig.~\ref{FourthLayerExample}, we illustrate the process of the GA-based channel parameter estimation, where $M_3 = 3$, $\big|\widetilde{y}_2^{(\widetilde{m})}|>|\widetilde{y}_1^{(\widetilde{m})}\big|$, and $\big|\widetilde{y}_3^{(\widetilde{m})}|>|\widetilde{y}_4^{(\widetilde{m})}\big|$. From \eqref{EqTheta} and \eqref{EqK},  $\Psi_{\rm L} = \widehat{\Omega}_{\rm n}$, $\Psi_{\rm R} = \widehat{\Omega}_{\rm n} + \overline{\Theta}_3$, $d_{\rm L} = \widehat{b}_{\rm n} - \overline{k}_3$, and $d_{\rm R} = \widehat{b}_{\rm n}$. Then the quantized samples can be obtained via \eqref{quantizedtk}. Note that one of the quantized samples is exactly the central codeword in the  hybrid-field neighboring search, whose received signal can be reused for the GA. Therefore, the training overhead of the GA-based channel parameter estimation is $M_3^2-1$.

The final results of the THBT are expressed as
\begin{align}
	\widehat{b}_{\rm f} =\widehat{b}_3,~\mbox{and}~\widehat{\Omega}_{\rm f} = \widehat{\Omega}_3.
\end{align}
According to \eqref{Definitionk}, the estimation of the distance can be expressed as 
\begin{align}\label{DistanceEstimation}
	\widehat{r}_{\rm f} = \frac{\lambda(1-\widehat{\Omega}_{\rm f}^2)}{4\widehat{b}_{\rm f}}.
\end{align}

Finally, we summarize the details of the GA-based third refinement method in \textbf{Algorithm~\ref{alg3}}.

Now we evaluate the computational complexity of the proposed THBT-ML and THBT-PSP. For the THBT-PSP, the first refinement method only  involves a straightforward  comparison  of the received signal power, resulting in a computational complexity of $\mathcal{O}(M_1)$. The second and third refinement methods provide closed-form expressions for channel parameter estimation, as shown in \eqref{Estimationofk} and \eqref{LSSolution}, respectively. Their computational complexities are $\mathcal{O}(M_2)$ and $\mathcal{O}(M_3^2)$, respectively. Therefore, the computational complexity of the  THBT-PSP  is $\mathcal{O}(\max\{M_1, M_2, M_3^2\})$. On the other hand, the THBT-ML shares the same first and third refinement methods as the THBT-PSP. However, during the second refinement step, it employs a two-dimensional search to solve \eqref{ML3}. Denote the number of the searches as $V$. Then, the computational complexity of the  THBT-ML  is $\mathcal{O}(V(2M_2+1))$.

\textbf{Remark 5:} The proposed scheme is also adaptable to the uniform planar array (UPA) configuration. According to Lemma 3 of \cite{JSAC23WZD}, the hybrid-field steering vector of a UPA can be approximated as the Kronecker product of channel steering vectors for the  horizontal uniform linear array (ULA) and the vertical ULA, which indicates that beam training of a UPA can be decoupled into the beam training of two ULAs. By performing the beam training for the horizontal ULA and the vertical ULA separately, the proposed scheme can be extended to the UPA configuration.

\begin{table*}[!t]
	\centering
	\renewcommand{\arraystretch}{1.2}
	\caption{Comparison of training overhead for different methods.}
	\label{tab1}  
	\begin{tabular}{cccc}
		\hline\hline\noalign{\smallskip}	
		\textbf{Methods} &  \textbf{Training Overhead}  &  \textbf{Parameter Settings} & \textbf{Calculated Training Overhead }\\
		\noalign{\smallskip}\hline\noalign{\smallskip}
		HFBS~\cite{ICCC2022CKJ} & $PQ$ & $P = 513,Q = 9$  & $4617$  \\
		TPBT~\cite{WCL22ZYP} & $N_{\rm t} + KQ$ & $N_{\rm t} = 513,K = 3,Q=9$ & 540  \\
		CHBT~\cite{TWC23SX} & $O+ 4(T-1)$ & $O = 33,T = 5$ & $49$    \\
		DHBT~\cite{CC22WXH} & $TR$ & $R = 100, T = 2$ & $200$   \\
		TSHBT~\cite{TVT23WCY} & $2W_1 + 4W_2$ & $W_1 =7$, $W_2 = 3$ & $26$\\
		Proposed THBT & $2(M_1 + M_2) + 3M_{\rm n} +M_3^2 + 3$ & $M_1=7,M_2=8,M_{\rm n}=3,M_3=2$ & 46 \\
		\noalign{\smallskip}\hline
	\end{tabular}
	\vspace{-0.1cm}
\end{table*}

\section{Simulation Results}\label{SimulationResults}
Now we evaluate the performance of the proposed THBT scheme. We consider an XL-MIMO system including a BS equipped with $N_{\rm t} = 513$ antennas and a single-antenna user. We set the wavelength as $\lambda = 0.005$~m corresponding to the carrier frequency of  $60$~GHz. The channel between the BS and the user is composed of one line-of-sight (LoS) path and two NLoS paths, where the channel gain of the LoS path obeys $g_1\sim\mathcal{CN}(0,\delta_1^2)$ and that of the NLoS paths obeys  $g_l\sim\mathcal{CN}(0,\delta_l^2)$ for $l\in\{2,3\}$. The channel angles   distribute   uniformly within $[-\sqrt{3}/2,\sqrt{3}/2]$. The HFBS~\cite{ICCC2022CKJ}, TPBT~\cite{WCL22ZYP},~CHBT~\cite{TWC23SX}, DHBT~\cite{CC22WXH}, and TSHBT~\cite{TVT23WCY} are adopted as benchmarks. The parameters of different methods are set in the \textbf{Parameter Settings} of Table~\ref{tab1}, where $P$, $Q$, $K$, $O$, $T$, $R$, $W_1$, and $W_2$ denote the angle samples, the distance samples, the number of candidate angles, the number of codewords in the first layer, the number of layers in the hierarchical codebook, and the number of space samples in each layer, the number of layers in the first stage of TSHBT, and the number of layers in the second stage of TSHBT, respectively. In addition, for the introduced parameters in the THBT, we set $\widetilde{k}_1 = -6.09\times10^{-5}$, $\overline{\Theta}_2 = 2/N_{\rm t}$, $\overline{k}_{\rm n} =  6/N_{\rm t}^2$ and $\overline{\Theta}_{\rm n} = 2/N_{\rm t}$.

\begin{figure}[!t]
	\begin{center}
		\includegraphics[width=80mm]{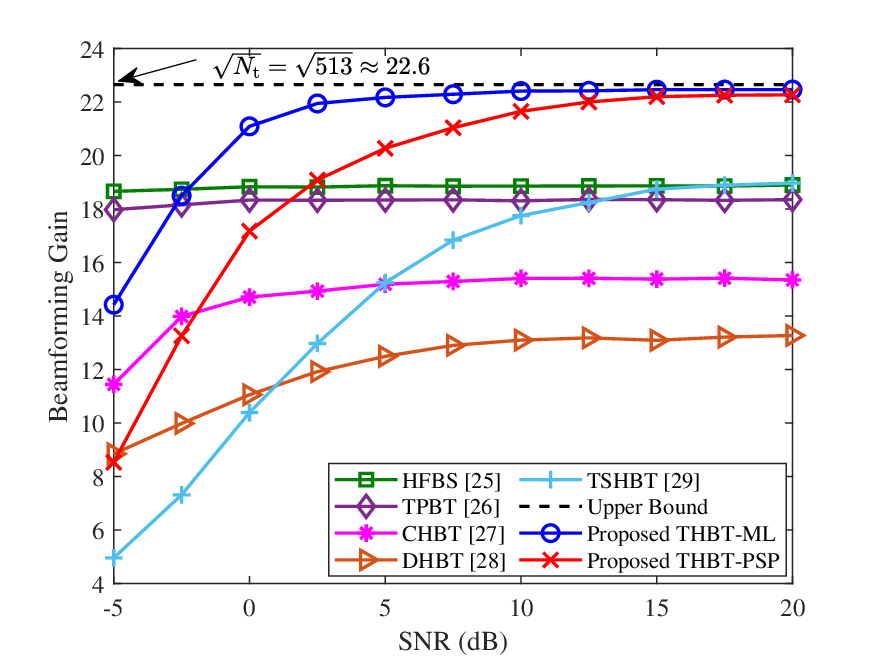}
	\end{center}
	\caption{Comparisons of the beamforming gains for different methods.	\label{BG}}
\end{figure}

In Fig.~\ref{BG}, we compare the beamforming gains of different methods. The beamforming gain after beam alignment is defined as 
	\begin{align}
		\xi  = \max_l\frac{|g_l|}{g_{\rm m}} |\boldsymbol{\alpha}(\Omega_l,r_l)^{\rm H}\boldsymbol{\alpha}(\Omega_{\rm f},r_{\rm f})|,
	\end{align} 
where $g_{\rm m} = \max_l |g_l|$. The distances between the BS and the user or scatterers obey the uniform distribution within $[10,200]$~m. We set $\delta_1 = 1$ and $\delta_{l} = 0.1$ for $l\in\{2,3\}$. From Fig.~\ref{BG}, at low signal-to-noise-ratios (SNRs), the HFBS achieves the best performance among all the methods; the justification is that the HFBS exhaustively tests the codewords in the hybrid-field codebook~\cite{ICCC2022CKJ} and needs far more times of beam training than other methods. Then the TPBT achieves the second-best performance due to  the accurate identification of candidate angles by far-field beam sweeping. With the increase of the SNR, the performance of the proposed THBT-ML and  THBT-PSP improves significantly and exceeds that of other methods when the SNR is larger than 5~dB. The advantage of the proposed methods mainly comes from the three steps of progressive refinement. In addition, the CHBT performs worse than other methods  because of the imperfect hierarchical codebook~\cite{TWC23SX} while the poor performance of the DHBT is attributed to the neglect of the polar-domain sparsity~\cite{Tcom22CMH}. The TSHBT performs worse than other methods at low SNRs due to low beamforming gains of the upper-layer codewords. Furthermore, at high SNRs, the proposed THBT-ML and THBT-PSP outperform other methods and can approach the upper bound.

\begin{figure}[!t]
	\begin{center}
		\includegraphics[width=80mm]{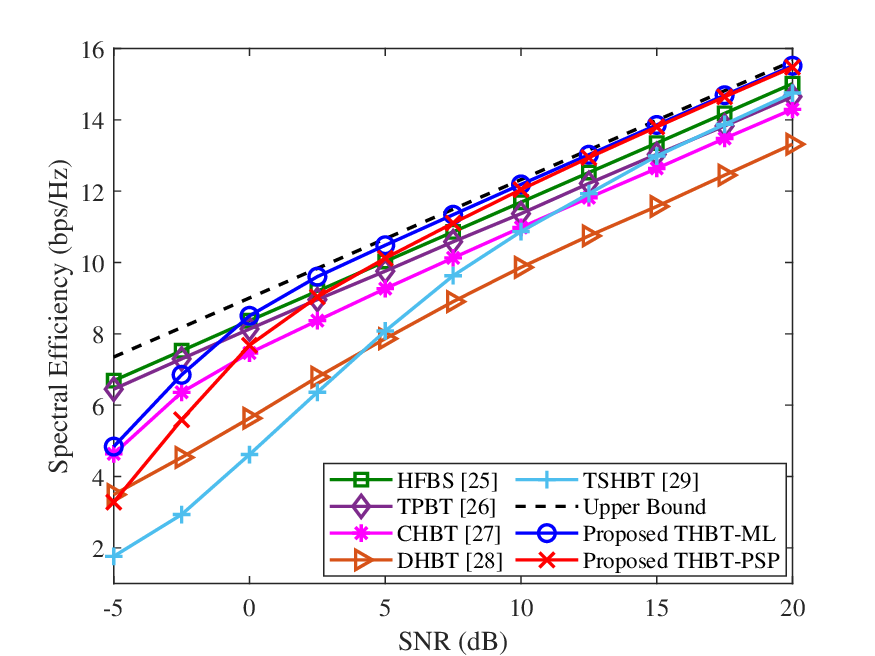}
	\end{center}
	\caption{Comparisons of the spectral efficiency for different methods. 	\label{SE}}

\end{figure}

In Fig.~\ref{SE},  we evaluate the spectral efficiency performance of different methods, where the simulation settings are the same as those in Fig.~\ref{BG}. From the figure, we can notice that the spectral efficiency performance is consistent with the beamforming gain performance in Fig.~\ref{BG}. In addition, the proposed THBT-ML and THBT-PSP can approach the upper bound with a negligible gap, e.g., 0.2 bps/Hz at $15$~dB. 

\begin{figure}[!t]
	\begin{center}
		\includegraphics[width=80mm]{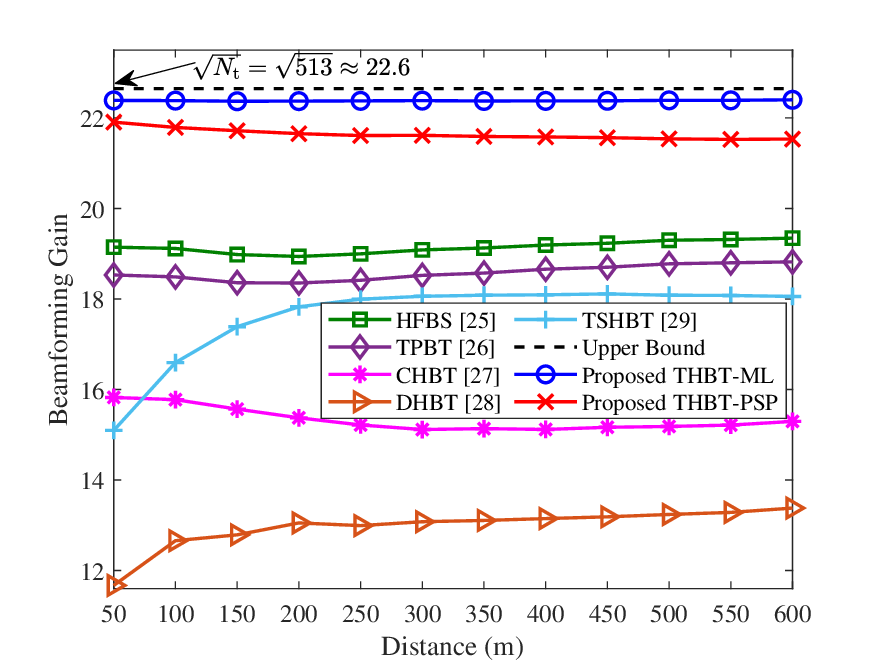}
	\end{center}
	 \caption{Comparisons of the beamforming gains for different methods  with varying distances.}	\label{ChangewithDistance}

\end{figure}

In Fig.~\ref{ChangewithDistance}, we compare the  beamforming gains of different methods with varying distances. The SNR is set to 10~dB and the distances between the BS and the user or scatterers obey the uniform distribution between $[10,r]$~${\rm m}$, where $r$ ranges from $50$~m to $600$~m. From the figure, the THBT-ML achieves the best performance among all the methods, followed by the THBT-PSP, and then the HFBS, TPBT, TSHBT, CHBT and DHBT in descending order. In addition, the THBT-ML, THBT-PSP, HFBS, and CHBT are robust to distance changes  due to the consideration of both the near-field and far-field effects. The performance of the TSHBT deteriorates at short distances because approximating the near-field channels with far-field ones by deactivating part of antennas may be not accurate enough for short distances. The performance of the DHBT deteriorates for short distances due to the neglect of the polar-domain sparsity.

\begin{figure}[!t]
	\begin{center}
		\includegraphics[width=80mm]{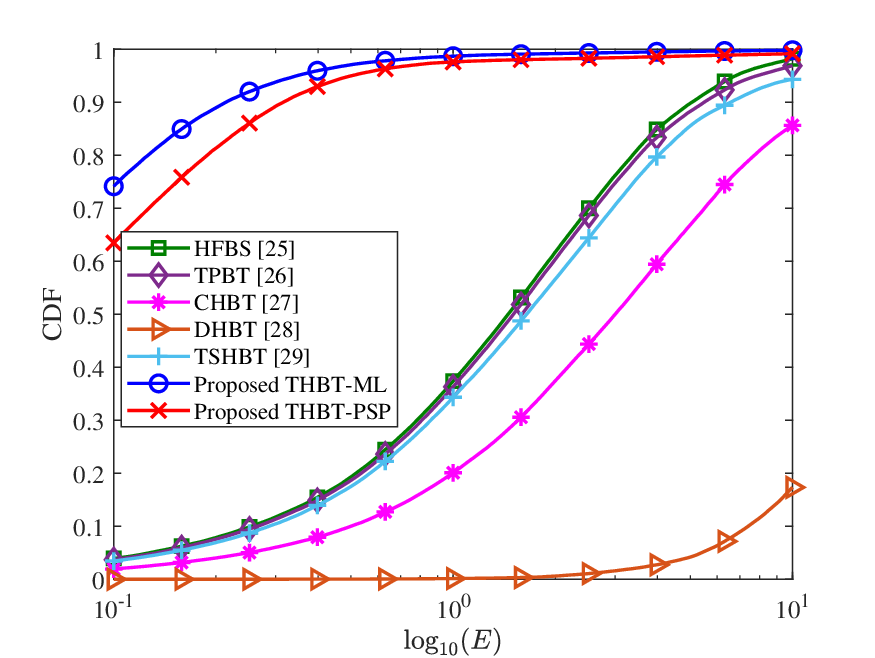}
	\end{center}
	\caption{Comparisons of the positioning performance for different methods.}\color{red}	\label{CDF}
\end{figure}

In Fig.~\ref{CDF}, we compare the positioning performance of different methods, where the SNR is set to 20~dB and the distances between the BS and the user or scatterers obey the uniform distribution between $[10,30]$~${\rm m}$. The deviation between the true position and the estimated position is denoted as $E$ and $10^5$ times of Monte Carlo simulation are used to calculate the cumulative distribution function (CDF). From the figure, the THBT-ML and THBT-PSP achieve much better performance than the other five methods. For example, when $E=1$~m, the values of CDF for  the  THBT-ML,  THBT-PSP, HFBS, TPBT, TSHBT, CHBT, and DHBT are $98.7\%$, $97.6\%$,~$37.4\%$,~$36.3\%$, $34.3\%$, $20.1\%$, and $0.13\%$, respectively, which indicates that the proposed schemes outperform existing ones and can attain high-accuracy positioning in most cases. 

\begin{figure}[!t]
	\begin{center}
		\includegraphics[width=80mm]{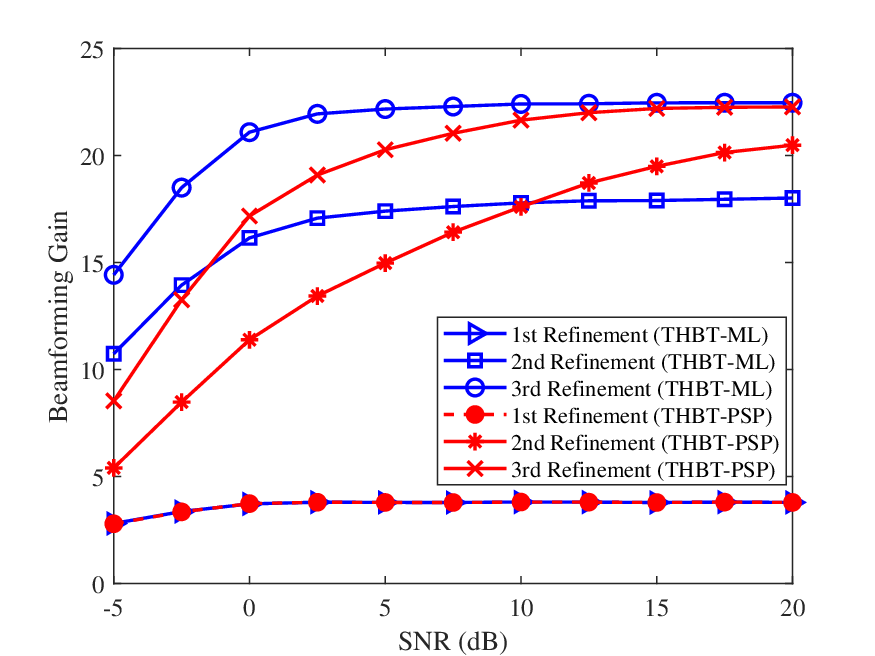}
	\end{center}
	\caption{Comparisons of the beamforming gains for different stages of refinement.}	\label{DL}
\end{figure}

\begin{figure}[!t]
	\begin{center}
		\includegraphics[width=82mm]{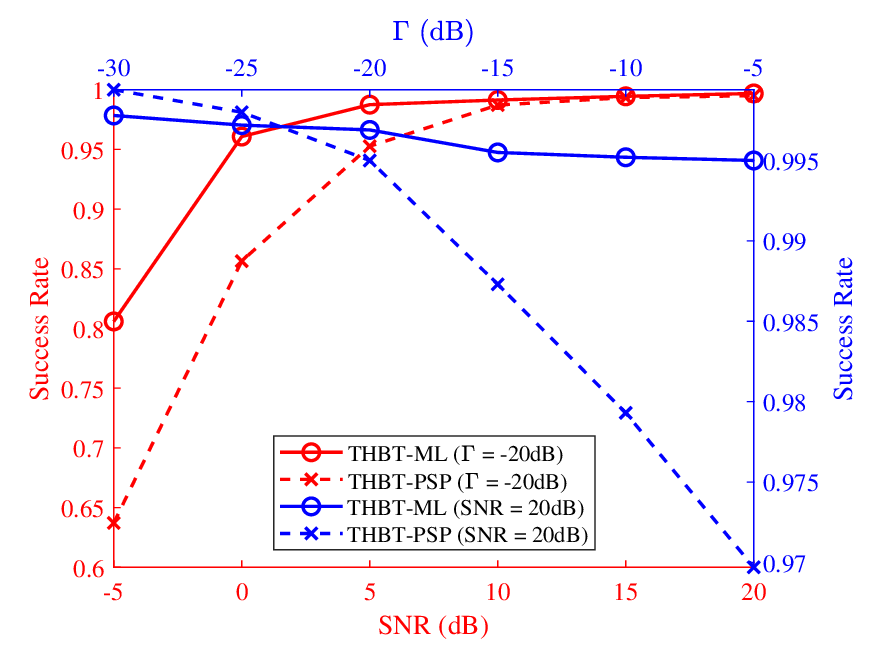}
	\end{center}
	 \caption{Illustration of the success rate of the neighboring search.}	\label{SR}

\end{figure}

In Fig.~\ref{DL}, we compare the beamforming gains of the THBT-ML and THBT-PSP for different stages of refinement. From the figure, the THBT-ML and THBT-PSP have similar performance for the first refinement, which is due to the fact that the THBT-ML and THBT-PSP share the same first-refinement beam training. In addition, the beamforming gains of both the THBT-ML and the THBT-PSP improve with the progress of the THBT. Moreover, the THBT-ML outperforms the THBT-PSP at low SNRs  due to the robustness of ML to noise while the performance of the THBT-PSP can approach that of the THBT-ML at high SNRs thanks to the exploitation of the phase property of the hybrid-field beam~gain.

In Fig.~\ref{SR}, we evaluate the performance of the neighboring search in terms of success rate, where $\Gamma$  denotes the ratio of the LoS path power to the NLoS path power, i.e., $\Gamma\triangleq 20\log_{10}(\delta_1/\delta_l)$ for $l\in\{2,3\}$. From the figure, the success rate  of the neighboring search increases with SNR and $\Gamma$, which indicates that the performance of the neighboring search is affected by the effects of noise and NLoS paths. In addition, when $\Gamma=-20$~dB and the SNR is larger than $5$~dB, the success rate of the neighboring search is considerably high, e.g. $98.8\%$ for the THBT-ML. Therefore, the neighboring search can successfully find the mainlobe of the channel path in most cases.

In Table~\ref{tab1}, we compare the training overheads of different methods. The training overheads of the HFBS, TPBT, CHBT, DHBT, TSHBT, and the proposed THBT are $PQ$, $N_{\rm t}+KQ$, $O+4(T-1)$, $TR$, $2W_1+4W_2$, and $2(M_1+M_2)+3M_{\rm n}+M_3^3+3$, respectively. Under the simulation settings, these six methods require $4617$, $540$, $49$, $200$, $26$, and  $46$ times of beam training, respectively. Specifically, when the SNR is larger than 5~dB, the proposed THBT outperforms the HFBS with a $99\%$ reduction in training overhead.

\section{Conclusion}\label{Conclusion}
In this paper, beam training for XL-MIMO systems has been investigated. By considering both the near field and far field, a triple-refined hybrid-field beam training scheme has been proposed, where the HFBG-based first refinement method, the  ML-based and PSP-based second refinement methods, and the GA-based third refinement method have been developed. Simulation results have shown that the proposed scheme outperforms the existing methods. In our future work, we will try to extend this work to the  THz band by considering the molecular absorption loss, THz spectrum windows, and beam split effects, following the works in \cite{Tcom22AI,Net23SA,WC23TJB}. In addition, we will also focus on reducing the attainable latency for beam alignment by jointly considering the training overhead and feedback.


\bibliographystyle{IEEEtran}
\bibliography{IEEEabrv,IEEEexample}

\end{document}